\documentclass[acmsmall,authorversion,nonacm]{acmart}
\usepackage{xcolor}
\usepackage{soul}
\usepackage{subcaption}
\usepackage{hyperref}
\usepackage{graphicx}
\usepackage{multirow}

\AtBeginDocument{%
  }

\setcopyright{cc}
\setcctype{by-nc-sa}
\acmDOI{10.1145/3748620}
\acmYear{2025}
\acmJournal{PACMHCI}
\acmVolume{9}
\acmNumber{6}
\acmArticle{GAMES025}
\acmMonth{10}
\received{2025-02-19}
\received[accepted]{2025-07-24}

\begin{document}


\title[First Contact with DPs in Chinese and Japanese Free-to-Play Mobile Games]{First Contact with Dark Patterns and Deceptive Designs in Chinese and Japanese Free-to-Play Mobile Games}

\author{Gloria Xiaodan Zhang}
\email{zhang.x.bj@m.titech.ac.jp}
\orcid{0009-0001-9648-8252}
\affiliation{%
  \institution{Institute of Science Tokyo}
  \city{Tokyo}
  \country{Japan}
}

\author{Yijia Wang}
\email{wang.y.cf@m.titech.ac.jp}
\orcid{0009-0004-2250-9163}
\affiliation{%
  \institution{Institute of Science Tokyo}
  \city{Tokyo}
  \country{Japan}
}

\author{Taro Leo Nakajima}
\email{nakajima.t.an@m.titech.ac.jp}
\orcid{0009-0006-7325-9074}
\affiliation{%
  \institution{Institute of Science Tokyo}
  \city{Tokyo}
  \country{Japan}
}
\affiliation{%
  \institution{Stuttgart Media University}
  \city{Stuttgart}
  \country{Germany}
}


\author{Katie Seaborn}
\orcid{0000-0002-7812-9096}
\affiliation{%
  \institution{Institute of Science Tokyo}
  \city{Tokyo}
  \country{Japan}
}
\affiliation{%
  \institution{University of Cambridge}
  \city{Cambridge}
  \country{UK}
}
\email{katie.seaborn@cst.cam.ac.uk}

\renewcommand{\shortauthors}{Zhang et al.}

\begin{abstract}
Mobile games have gained immense popularity due to their accessibility, allowing people to play anywhere, anytime. Dark patterns and deceptive designs (DPs) have been found in these and other gaming platforms within certain cultural contexts. 
Here, we explored 
DPs in the onboarding experiences of free-to-play mobile games from 
China and Japan. We identified several unique patterns 
and mapped their relative prevalence. We also found that game developers often employ combinations of DPs as a strategy (``DP Combos'') and use elements that, while not inherently manipulative, can enhance the impact of known patterns (``DP Enhancers''). Guided by these findings, we then developed an enriched ontology for categorizing deceptive game design patterns into classes and subclasses. This research contributes to understanding deceptive game design patterns 
and offers insights for future studies on cultural dimensions and ethical game design in general.
\end{abstract}




\maketitle

\section{Introduction}
\label{sec:intro}
The rise of video games has sparked widespread discussion regarding their use in various domains~\cite{kato2010video, de2003video}. Games are considered to offer numerous benefits to players beyond mere entertainment and escape~\cite{granic2014benefits}, including cognitive improvements~\cite{sennersten2009investigation}, 
motivational benefits~\cite{sweetser2005gameflow, przybylski2010motivational}, 
emotional benefits~\cite{russoniello2009eeg, vuorre2024affective}, 
and social benefits~\cite{gentile2009effects, granic2014benefits}. 
However, concerns have also emerged regarding potential negative effects, such as problematic addiction and gambling-related issues~\cite{bean2017video, zendle2020beyond}.
Recent work has focused on specific problematic game design elements, such as loot boxes~\cite{zendle2020beyond, xiao2022probability, neely2021come}. In the field of Human-Computer Interaction (HCI), these manipulative and disruptive designs are referred to as ``dark patterns'' or ``deceptive designs'' (DPs)\footnote{In line with \href{https://www.acm.org/diversity-inclusion/words-matter}{ACM}, we avoid using the term ``dark patterns'' because ``dark'' is unclear (it does not mean colour, despite being a UI matter) and may reinforce negative associations as a term used to describe people.}. DPs are UI/UX elements that undermine, impair, or distort the user's ability to make autonomous and informed decisions when interacting with digital systems, irrespective of the designer’s intent~\cite{gray2024ontology, zhang-kennedy2024navigating, chivukula2018darkint, sanchez2023ethical}. DPs have been examined in various contexts, including social media~\cite{schaffner2022understanding, karagoel2021dark}, e-commerce~\cite{mathur2019dark}, mobile applications~\cite{di2020ui}, and gaming~\cite{zagal2013dark}. 
DPs in games are distinct from general UI/UX DPs as they are deeply rooted in gameplay, leading researchers to focus on game DPs over general UI patterns~\cite{zagal2013dark, sousa2023dark, dahlan2022finding, hadan2024, king2023investigating, fitton2019creating}.

Most research on DPs in games~\cite{hadan2024, king2023investigating, sousa2023dark, dahlan2022finding} has focused on evaluating from either expert or player perspectives using the game DP classifications by \citet{zagal2013dark}. While general UI/UX DPs have been systematically organized into structured ontologies, game DPs remain largely unstructured and fragmented within the current literature~\cite{sousa2023dark, dahlan2022finding, zagal2013dark}. Additionally, there is a notable absence of exploratory and descriptive analyses of game DPs~\cite{sousa2023dark, dahlan2022finding} and a missing cultural angle~\cite{hidaka2023linguistic}. 
Empirical investigations remain limited, particularly in examining game DPs across diverse game genres and business models~\cite{dahlan2022finding, hadan2024comba}. To address these gaps, we extended previous studies~\cite{hidaka2023linguistic, sousa2023dark, dahlan2022finding} to explore DPs in the context of mobile games in Japan and China. Since mobile games and especially free-to-play apps are popular in these countries~\cite{jitsuzumi2022policies}, we focused on these offerings. We selected different genres of games to discover and gain a deeper understanding of DPs in game settings. Both China and Japan rank among the top five global mobile game markets by revenue\footnote{\url{https://www.statista.com/outlook/dmo/digital-media/video-games/mobile-games/worldwide}}, and their game industries have maintained close ties for decades, influencing each other significantly~\cite{nakamura2024japanese, chen2014s, wang2024globalization}. This establishes a solid foundation for investigating DPs in contemporary game markets and examining how they are implemented within real-world industry practices. 

In this exploratory and descriptive work, we aimed to provide evidence of established DPs in free-to-play Chinese and Japanese mobile games, uncover new game DPs, and examine how DPs work together and are shaped by other elements of the gameplay and game design. We scoped our work to the onboarding phase, a key stage in free-to-play apps of all kinds~\cite{cheung2014firsthour}, aiming to examine how DPs target and ``hook'' new players. We asked several complementary research questions (RQs): \textit{(RQ1) What kinds of DPs are present in Japanese and Chinese free-to-play mobile games, and how do they interplay? (RQ2) What is the distribution of each pattern? (RQ3) What are the similarities and differences in types, frequencies, and aesthetics between the Japanese and Chinese offerings?} 
To answer these questions, we adopted and refined the qualitative heuristic analysis methodology used by \citet{hidaka2023linguistic, sousa2023dark, dahlan2022finding}, selecting 18 mobile games in total (nine from the Japanese App Store and nine from the Chinese App Store) and providing thick descriptions~\cite{Ponterotto_2015thick} of the DPs found. We offer four main contributions:

\begin{itemize}
    \item \textbf{Ontology Extension:} We extended the existing game-oriented frameworks by \citet{zagal2013dark}, \citet{sousa2023dark}, and \citet{king2023investigating} into a consistent three-level hierarchical ontology, drawing from the general framework by \citet{gray2024ontology} and a lay resource: ``the Dark Pattern Games'' website\footnote{\url{https://www.darkpattern.games}}. 
    \item \textbf{Evidence of DPs in Free-to-Play Mobile Games}: We offer empirical findings on the presence and distribution of DPs in contemporary popular mobile games, with comparisons to previous studies and between our national samples.
    \item \textbf{Identification of New Patterns:} We discovered several new DPs, ranging from the high-level Technical pattern to specific low-level patterns, like Overloading.
    \item \textbf{Discovery of DP Combos and Enhancers:} We identified novel categories of designs that illustrate how, when, and where DPs are implemented in gameplay. Notably, DP Enhancers are elements that, while not inherently malicious, operate within the game context to enhance the impact of DPs.
\end{itemize}

Our work fills existing gaps in DP research by providing descriptive and exploratory insights into how these patterns can manifest in free-to-play mobile games and setting the stage for sociocultural and cross-cultural work. This work aims to benefit scholars, the game industry, and individual players by deepening our understanding and awareness of DPs in games.

\section{Background}
\label{sec:background}
Technological advancements and especially the mobile phone industry have transformed modern gaming. Rather than being solely a product, such as a cartridge, games are now often characterized by a model known as ``Game-as-a-Service'' (GaaS), where the experience is supplemented by ongoing services~\cite{wilhelmsson2022shift}. 
The transition from a single transaction to long-term relationships between players and developers allows companies to offer games for free while generating revenue through \emph{microtransactions}~\cite{zendle2020beyond,neely2021come}, such as in-game purchases. This \emph{free-to-play} model, a staple of modern mobile gaming~\cite{neely2021come,hadan2024}, has emerged as one of the most popular monetization strategies, providing companies with competitive market advantages while enabling players to access the game at no initial cost~\cite{massarczyk2019economic}. Microtransactions have the potential to generate indefinite revenue for game developers. This is facilitated by in-game purchasable content that can sustain ongoing sales, contingent upon sustained player engagement with the game~\cite{kostic2021implementation}. Meanwhile, major platforms like the App Store (Apple), Google Play, and Microsoft Store classify ``Games'' as a primary category, with separate rankings for ``Free-to-Play'' and ``Buy-to-Play'' games (those requiring purchase for access). These always-available online stores facilitate quicker transactions and reduce interruptions to gameplay, often prompting players to exceed their initial spending intentions~\cite{gibson2023videogame, Petrovskaya2022micro}. These attributes raise the question of deceptive design.

Developers can strategically design in-game content to influence and motivate player purchasing behaviour~\cite{hamari2017players}. This can be deceptive and manipulative, potentially leading to negative gaming experiences for players~\cite{hadan2024}. On this front, microtransactions, including loot boxes, battle passes, and pay-to-win mechanisms, have been extensively discussed~\cite{zendle2020prevalence, lelonek2021pay, joseph2021battle}. \emph{Loot boxes} (also ``loot crates,'' ``loot cases,'' and ``loot chests'')~\cite{yokomitsu2021characteristics,li2019relationship}, provide virtual items as rewards each time they are opened. 
The element of randomness is controversial due to its similarity to gambling, where chance-based outcomes can lead to addiction~\cite{yokomitsu2021characteristics}. A \emph{battle pass} (also called ``seasonal'' or ``event'' passes) is an in-game purchasable item that grants the ability to earn additional rewards through continued gameplay. Battle passes can intersect with other monetization strategies, such as pay-to-win and loot boxes~\cite{nieborg2015crushing, joseph2021battle}. Players risk receiving no reward without continuously investing time in the game, even if they have used real-world money. \emph{Pay-to-win} microtransactions refer to any in-game content that players can purchase with real-world money to increase their chances of success~\cite{zendle2020changing}. This is directly tied to functional aspects that assist players in achieving their objectives, such as conquering dungeons, defeating monsters, and prevailing over other players, such that it can be perceived as cheating~\cite{freeman2022ingamepurch}.
Microtransactions remain tricky. 
Although countries like Belgium have banned loot boxes by threatening criminal prosecution of game companies without a gambling license, 
players can easily circumvent the ban~\cite{xiao2023breaking}. 

Narrowing the scope to DPs in games, \citet{zagal2013dark} introduced the concept of ``dark game design patterns,'' emphasizing that these designs are intentionally created to cause negative player experiences and benefit game developers. They identified three main classes---Temporal, Monetary, and Social Capital-based patterns---and acknowledged a ``grey'' area for some patterns, where the effects could be contingent on the player's personal situation. The \emph{Dark Pattern Games} website\footnote{\url{https://www.darkpattern.games}} was established to encourage visitors to educate themselves and explore ``healthy games'' versus ``dark games.'' \citet{king2023investigating} investigated player perception of deceptive elements designed to promote monetization in free-to-play 3D games. \citet{hadan2024} explored player perceptions of deceptive designs in \emph{Overwatch}, highlighting the game's transition from a buy-to-play to a free-to-play business model. In the mobile game space, previous studies have concentrated on a specific casual genre~\cite{dahlan2022finding} and those aimed at young children~\cite{sousa2023dark}. 
Hence, we extend this body of work by focusing on the broader cases of mobile games that target a general group of players. 

An initial sustained play session or \emph{onboarding experience} is a key part of the player journey, serving as the entrance to the primary gaming experience~\cite{cheung2014firsthour}. This is particularly critical for free-to-play mobile games: the player incurs no cost and the company does not benefit when the player is dissatisfied by their ``first contact'' with the game and decides to switch to another option. Since the games are free, a player can easily opt for alternatives in the app store. Given this critical window for capturing attention, we focused on DPs targeting players during the initial phase of gameplay. If players are not ``hooked'' early on, they have little incentive to continue playing~\cite{cheung2014firsthour}.

\section{Methodology}
\label{sec:methods}
We employed the \emph{heuristic} or qualitative expert classification approach from previous research~\cite{di2020ui, hidaka2023linguistic}, which involves analyzing the entire interactive experience rather than relying on static screenshots. The methodology was divided into two main sections: \emph{recording} and \emph{classification}. Three researchers with backgrounds in HCI and games participated. Modifications were made to accommodate constraints such as time, budget, and available devices, as well as to adapt to the specific characteristics of mobile gaming~\cite{dahlan2022finding}. Prior to analysis, all underwent an hour of training on DPs. All three were experienced gamers, each having logged at least 2,000 hours across various genres of PC and mobile games. Additionally, one researcher had app development experience. Two were native Chinese speakers with fluent Japanese skills, enabling fluent playing sessions. Our protocol was registered on OSF\footnote{\url{https://osf.io/tv6a9}} on May 27\textsuperscript{th}, 2024, prior to recording.

\subsection{Sample Selection}
Building on \citet{dahlan2022finding}, we aimed to analyze the most successful and popular mobile games on the market. Prior research has found a positive association between offering free apps and improved sales performance~\cite{lee2014determinants}. Therefore, on May 22\textsuperscript{nd}, 2024, we collected the top 50 games in the ``Top Free Games'' category from the App Store (Apple) for each nation. We used the classification system by Nihon Kogakuin\footnote{\url{https://www.neec.ac.jp/department/design/gamecreator/type}}. Genres included: shooting games, action games, role-playing games, adventure games, racing games, puzzle games, simulation games, sandbox games, music games, and table games. Not all mobile game developers categorized their games by genre on the App Store. Notably, sandbox games were not categorized. Since it is difficult to determine whether a game qualifies as a sandbox without firsthand play experience (based on the definition of sandbox games as lacking specific goals\footnote{\url{https://www.neec.ac.jp/department/design/gamecreator/type}}), sandbox games were excluded. 
We aimed to ensure each set represented the nine genres. We ultimately chose 18 games (nine from each nation), in line with the sample sizes in previous qualitative studies (ranging from three to 29 games)~\cite{Bhatnagar2024, hadan2024comba, Jonas2019, korhonen2006, hochleitner2015heuristic}. 

Games were selected based on two criteria: (1) ranking, from top to bottom, and (2) publication date, i.e., available for over six months. Although the App Store did not disclose what determined app rankings, \citet{karagkiozidou2019app} summarized criteria that significantly influence these rankings, such as install volume, ratings, and user reviews. This rationale underpinned our decision to select top-ranked games. 
Game publishers often provide pre-download versions of games before official launch, causing a rapid rise in the ``Top Free Games'' rankings. For accuracy, stability, and avoidance of novelty effects~\cite{jiao2022empirical}, 
we chose games that had been published for over six months.


\subsection{Recording Section}
Three researchers participated in the recording phase. Given the onboarding context, each researcher created a new local Apple ID and downloaded the selected games from the target App Stores, regardless of prior gameplay experience. Since all games were available on iOS, the researchers used iPhone devices. 
The recording began before launching the game app. 
Researchers created new characters, and did not skip any part of the game, including the tutorial session.

A proposed rule of thumb is a 5-10 minute onboarding time for games~\cite{Bycer_2019}. Expert evaluations of player UX have also used a duration of 30 minutes~\cite{hochleitner2015heuristic}. To ensure a continuous gaming experience, each researcher played every game at least two times for at least 15 minutes each time. All three researchers maintained a consistent pace, even though only one play session was recorded for analysis. We focused on first-hand, repeated gaming experiences to gain a comprehensive understanding of the game systems and a multifaceted perspective on the games as both artefacts and dynamic activities~\cite{lankoski2015formal}. We aimed to cover the entire onboarding experience, so each researcher freely extended the duration of each session, as needed. Every available interface and component in every game was accessed and carefully examined in addition to ordinary gameplay. Event interfaces, in-game stores, and other areas were thoroughly reviewed. However, not all onboarding DPs may have been captured, as gameplay is continuous and path-dependent. Once a researcher chose a pathway, alternative routes---and any potential DPs---were missed.

\subsection{Classification Section}
Next, the researchers formed pairs and collaboratively reviewed the game recordings. This process reduced the likelihood of DP evasiveness\footnote{We avoid the term ``DP blindness,'' as use of blindness as a metaphor for failure is ableist and \href{https://www.acm.org/diversity-inclusion/words-matter}{words matter}.} and allowed for consensus-driven decisions~\cite{di2020ui,hidaka2023linguistic}. Pairs worked together to identify and document the DPs with timestamps. In cases of no consensus, a third researcher participated. Decisions were made through majority voting. If the identified DPs exhibited the same design and appearance, i.e., same case, they were counted once. However, if the same DPs occurred in different scenarios, such as loot boxes containing aesthetic items (e.g., fashion) and another set of loot boxes containing functional items (e.g., weapons and armour), they were treated as distinct cases.

\subsection{Taxonomy and Data Analysis}
\citet{zagal2013dark} developed categories of ``dark game design patterns'' by integrating insights from developers, their personal experiences as gamers, and critical feedback from other players. The categories were: Temporal DPs, Monetary DPs, and Social Capital-Based DPs. 
\citet{sousa2023dark} expanded this framework by incorporating ``Psychological DPs,'' drawing on insights from the \emph{Dark Pattern Games} website\footnote{\url{https://www.darkpattern.games}}. In kind, we used these established categories~\cite{zagal2013dark, sousa2023dark, gray2024ontology} as a codebook and then sorted cases into various subcategories. Since mobile games are a type of app, we also employed the general DP categories from \citet{gray2024ontology}. We thereby extended the existing taxonomy for games by identifying new classes and subclasses that better fit contemporary online gaming, which became a contribution of our work.
We provide a concise overview of the main categories (with more details in \autoref{sec:findings} and \autoref{sec:discussion}):

\begin{itemize}
    \item \textbf{Temporal DPs}
 manipulate the amount of time spent on gameplay, 
i.e., expected versus actual time,
leading players to feel that their time has been unfairly exploited~\cite{zagal2013dark}.

\item \textbf{Monetary DPs}
 cause players to unexpectedly invest more financial resources, ultimately causing them to regret or lose track of their expenditures~\cite{zagal2013dark}.

\item \textbf{Social DPs}
 exploit players' social connections from the real world and in-game to deceive or manipulate them, thereby putting their social standing and relationships at risk~\cite{zagal2013dark, sousa2023dark}.

\item \textbf{Psychological DPs}
 and practices manipulate players' minds, emotions, and cognition, inducing them to make poor decisions through psychological tricks~\cite{sousa2023dark}.

\end{itemize}

We provide counts, percentages, and, where possible, mean (M) and standard deviation (SD). 

\section{Findings}
\label{sec:findings}
\begin{figure}
\centering
\begin{subfigure}{.48\textwidth}
  \centering
  \includegraphics[width=1\linewidth]{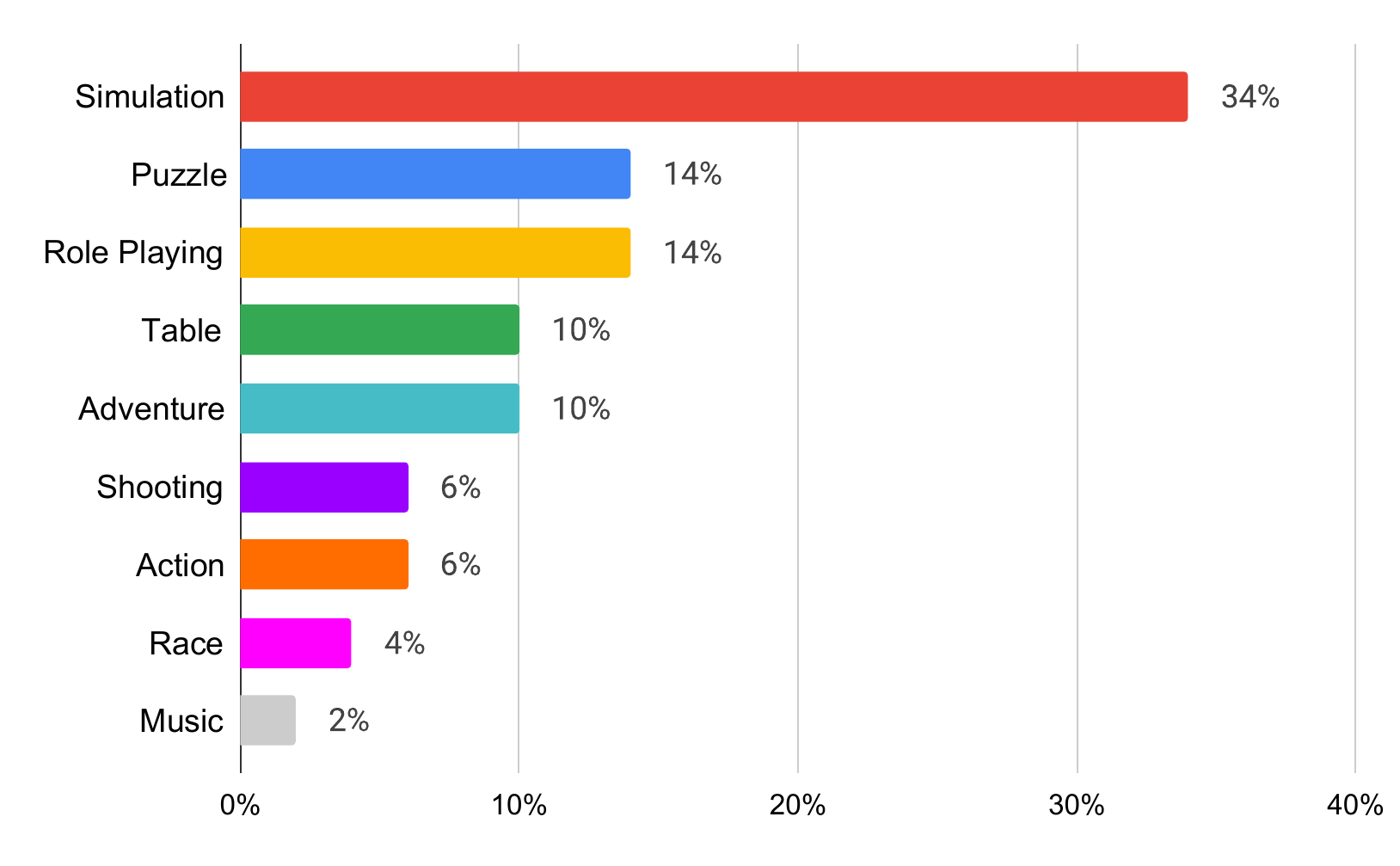}
  \caption{Game genres in the Chinese App Store.}
  \label{fig:Game Genres in Chinese App Store}
\end{subfigure}%
\begin{subfigure}{.48\textwidth}
  \centering
  \includegraphics[width=1\linewidth]{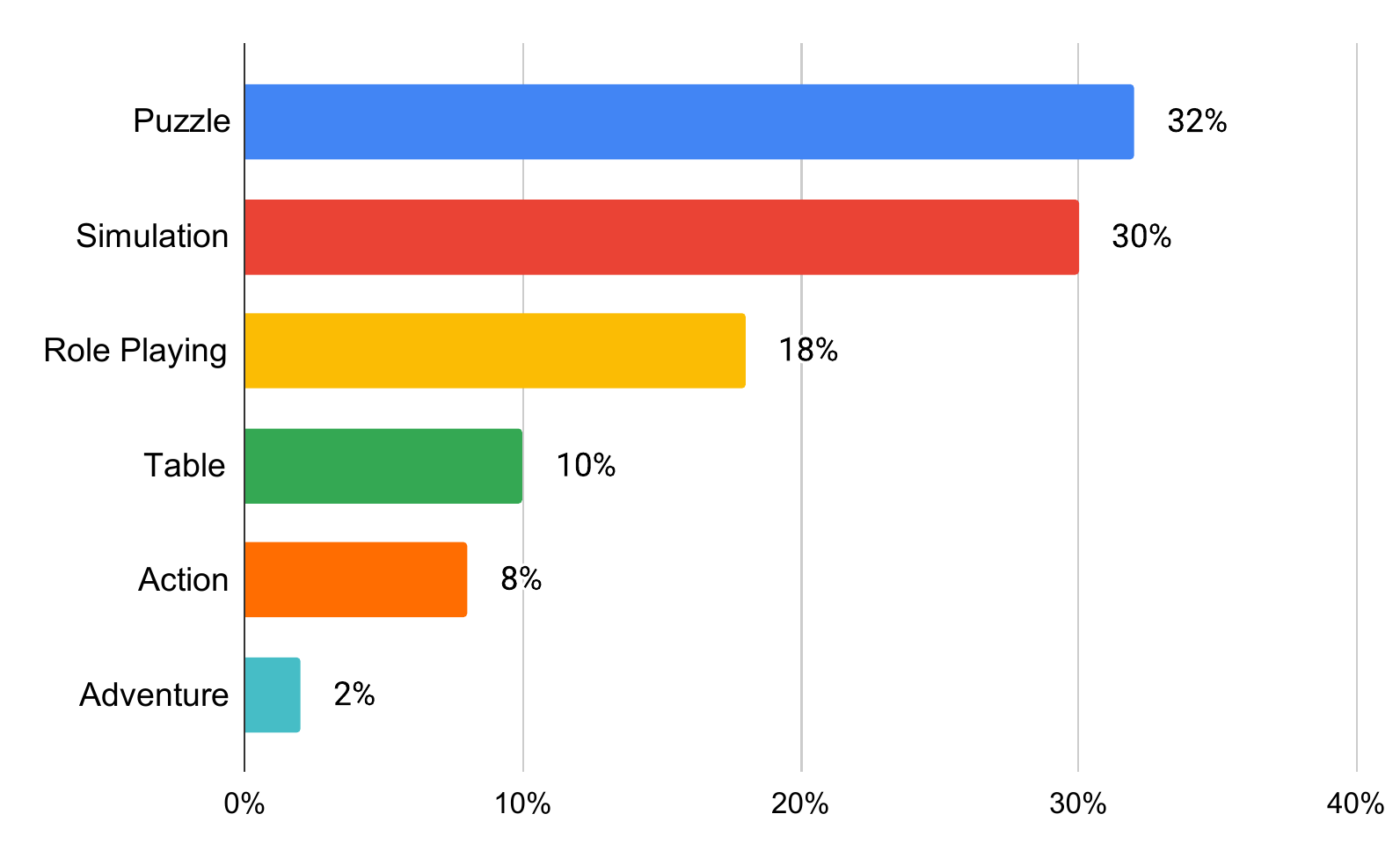}
  \caption{Game genres in the Japanese App Store.}
  \label{fig:Game Genres in Japanese App Store}
\end{subfigure}
\caption{Game genres rankings in top 50 free games for the Chinese (a) and Japanese (b) App Stores.}
\Description{Two figures illustrating the ranking of the top 50 free games by category in the Chinese and Japanese App Stores. Simulation, Puzzle, and Roleplaying are the top options across both stores.}
\label{fig:genres rankings}
\end{figure}

The top 50 free games on the Japanese and Chinese App Stores revealed dominant genres, as illustrated in \autoref{fig:genres rankings}. 
The corpus was selected based on a list of games that met the two predefined criteria (\autoref{sec:methods}). However, shooting, music, and racing games were absent from the top 50 rankings on the Japanese App Store. Thus, we reviewed the extended rankings
. The length of each recording session is in \autoref{table:durec}\footnote{The institutional ethics board did not allow disclosure of game names, so we use location and genre instead.}. We provide our anonymous data set online\footnote{\url{https://bit.ly/firstcontactgames}}.

\begin{table}[!ht]
\caption{Time duration of each recording session; recordings may be separated due to technical issues.}
\label{table:durec}
\begin{tabular}
{p{0.21\linewidth}p{0.3\linewidth}p{0.3\linewidth}}
\toprule
\textbf{Sample Game} &
  \textbf{Session 1} &
  \textbf{Session 2} 
   \\ \midrule
%
%
%
\textbf{CN: Simulation} &
  \text{17min 21sec} &
  \text{26min 01sec}
   \\

\textbf{CN: Table} &
  \text{53min 24sec} &
  \text{21min 28sec}
   \\

\textbf{CN: Role Playing} &
  \text{04min 32sec \& 13min 34sec } &
  \text{43min 05sec}
   \\

\textbf{CN: Puzzle} &
  \text{19min 33sec} &
  \text{10min 11sec \& 07min 29sec}
   \\

\textbf{CN: Adventure} &
  \text{18min 03sec} &
  \text{57min 25sec}
   \\

\textbf{CN: Action} &
  \text{15min 02sec} &
  \text{20min 41sec}
   \\

\textbf{CN: Music} &
  \text{15min 45sec} &
  \text{24min 28sec}
   \\

\textbf{CN: Race} &
  \text{16min 10sec} &
  \text{23min 03sec}
   \\

\textbf{CN: Shooting} &
  \text{35min 58sec} &
  \text{30min 14sec}
   \\

\textbf{JP: Simulation} &
  \text{20min 05sec} &
  \text{20min 57sec}
   \\

\textbf{JP: Table} &
  \text{32min 08sec} &
  \text{17min 23sec}
   \\

\textbf{JP: Role Playing} &
  \text{11min 18sec \& 10min 17sec} &
  \text{17min 54sec}
   \\

\textbf{JP: Puzzle} &
  \text{19min 58sec} &
  \text{15min 19sec}
   \\

\textbf{JP: Adventure} &
  \text{15min 36sec} &
  \text{16min 07sec}
   \\

\textbf{JP: Action} &
  \text{15min 59sec} &
  \text{18min 13sec}
   \\

\textbf{JP: Race} &
  \text{15min 07sec} &
  \text{14min 59sec}
   \\

\textbf{JP: Music} &
  \text{16min 18sec} &
  \text{15min 42sec}
   \\

\textbf{JP: Shooting} &
  \text{18min 47sec} &
  \text{19min 08sec}
   \\ \bottomrule
   \end{tabular}
\end{table}

\subsection{DPs in Chinese and Japanese Mobile Games (RQ1)}

We 
identified gaps in the existing game DP taxonomies. We observed that certain DPs relied on other DPs, i.e., DP Combos, and discovered non-DP design patterns that were co-present with and boosted a given DP
, i.e., DP Enhancers. We therefore integrated concepts from the game side~\cite{zagal2013dark,sousa2023dark}\footnote{\url{https://www.darkpattern.games}} to enrich the game branch of DP ontologies, following the basic structure of \citet{gray2024ontology}. \textbf{High-level patterns} represent abstract and general strategies, including Temporal, Monetary, Social, Psychological, and the newly discovered Technical patterns. \textbf{Meso-level patterns} address the specific ``angle of attack,'' while \textbf{low-level patterns} represent the most detailed and context-specific forms of patterns. 
The novel patterns are highlighted in \hl{yellow} in \autoref{table:fradp}. All known game DPs documented on the ``Dark Pattern Games'' website\footnote{\url{https://www.darkpattern.games}} and in prior literature~\cite{zagal2013dark, sousa2023dark, king2023investigating, hadan2024, dahlan2022finding} were included in our analysis. Following \citet{gray2024ontology}, we first compiled a comprehensive list of all game DPs. We then organized these patterns into hierarchical levels based on direct citations and inferred relationships. Meso-level categories were created and named to systematically integrate both previously identified and newly discovered game DPs.


\begin{table}[!t]
\caption{Enriched ontology of DPs in games, extending the taxonomies of \citet{zagal2013dark}, \citet{sousa2023dark}, \citet{king2023investigating}, \citet{hadan2024}, and the ``Dark Pattern Games website,'' listed by type. Novel items and updates are presented in \hl{yellow}.}
\label{table:fradp}
\resizebox{\textwidth}{!}{%
\begin{tabular}{p{0.22\linewidth}p{0.48\linewidth}p{0.57\linewidth}}
\toprule
\textbf{High-Level} & \textbf{Meso-Level} & \textbf{Low-Level} \\ \midrule

\multirow{6}{*}{\parbox{\linewidth}{\raggedright\textbf{Temporal \\ \cite{zagal2013dark, sousa2023dark, dahlan2022finding, hadan2024}}}} &
  \textbf{Playing by Appointment~\cite{zagal2013dark, sousa2023dark, dahlan2022finding, hadan2024}} &
  \emph{Evil} Battle Pass~\cite{hadan2024} $\bullet$ Wait to Play \\
 &
  \textbf{Daily Rewards~\cite{hadan2024, sousa2023dark}} &
   \\
 &
  \textbf{Grinding~\cite{zagal2013dark, sousa2023dark, dahlan2022finding, hadan2024}} &
   \\
 &
  \textbf{Advertisement~\cite{sousa2023dark}} & 
   \\
 &
  \textbf{Infinite Treadmill~\cite{hadan2024}} &
   \\
 &
  \textbf{Mandatory Marathon} &
  Can't Pause or Save \\
  
\midrule
\multirow{13}{*}{\parbox{\linewidth}{\raggedright\textbf{Monetary~\cite{zagal2013dark}}}} &
  \textbf{Pay to Progress} &
  Pay to Skip~\cite{zagal2013dark, sousa2023dark, dahlan2022finding, hadan2024} $\bullet$ Pay to Win~\cite{zagal2013dark, dahlan2022finding} $\bullet$ Pay Wall~\cite{sousa2023dark} $\bullet$ Narrative Obligation~\cite{king2023investigating} \\
 &
  \textbf{Intermediate Currency} &
  Premium Currency~\cite{king2023investigating, sousa2023dark, hadan2024} $\bullet$ \hl{Polymorphic Currency} $\bullet$ Bulk Purchase~\cite{hadan2024} $\bullet$ Leftovers~\cite{hadan2024} \\
 &
  \textbf{Deceptive Luxury} &
  Artificial Scarcity~\cite{hadan2024} $\bullet$ \hl{Remedy Consumption} \\
 &
  \textbf{Recurring Fee} & \\
 &
  \textbf{Gambling} &
  Loot Boxes~\cite{hadan2024, sousa2023dark} $\bullet$ \hl{Risk Conversion} $\bullet$ \hl{Bundle Bonus} \\
 &
  \textbf{Power Creep} & \\
 &
  \textbf{Waste Aversion~\cite{hadan2024}} & \\
 &
  \textbf{Easy to Purchase} &
  Anchoring Tricks~\cite{hadan2024} $\bullet$ Accidental Purchases~\cite{king2023investigating} $\bullet$ Low barrier~\cite{king2023investigating} $\bullet$ Prompted to buy~\cite{king2023investigating} \\
 &
  \textbf{UI Misdirection~\cite{king2023investigating}} &
  Exciting UI Effect~\cite{king2023investigating} \\
 &
  \textbf{Never-Ending Lure} &
  \hl{First Charge Discount} $\bullet$ \hl{Accumulating Rewards} \\

\midrule
\multirow{7}{*}{\parbox{\linewidth}{\raggedright\textbf{Social~\cite{zagal2013dark}}}} &
  \textbf{Forced Fellowship} &
  Social Pyramid Scheme~\cite{zagal2013dark, dahlan2022finding, sousa2023dark} $\bullet$ Social Obligation (Guilds)~\cite{sousa2023dark} \\
 &
  \textbf{Friend Spam  $\bullet$ Impersonation~\cite{zagal2013dark, dahlan2022finding, sousa2023dark}} & \\
 &
  \textbf{Reciprocity} & \\
 &
  \textbf{Encourages Anti-Social Behavior} & \\
 &
  \textbf{Fear of Missing Out (FOMO)~\cite{king2023investigating, hadan2024, sousa2023dark}} & \\
 &
  \textbf{Competition~\cite{zagal2013dark}} & \\

\midrule
\multirow{6}{*}{\parbox{\linewidth}{\raggedright\textbf{Psychological~\cite{sousa2023dark}}}} &
  \textbf{Easy to Get, Hard to Lose} &
  Invested (Endowed) Value~\cite{sousa2023dark} $\bullet$ Endowed Progress~\cite{hadan2024} \\
 &
  \textbf{Complete the Collection~\cite{sousa2023dark}} & \\
 &
  \textbf{Illusion of Control} & \\
 &
  \textbf{Aesthetic Manipulation~\cite{sousa2023dark, hadan2024, king2023investigating}} & \\
 &
  \textbf{Optimism and Frequency Biases} & \\
 &
  \textbf{Reward Mania} &
  \hl{Overloading} $\bullet$ \hl{Sycophant} $\bullet$ Variable Rewards~\cite{sousa2023dark} \\

\midrule
\textbf{\hl{Technical}} &
  \textbf{\hl{Fragmented Downloads}} & \\

\bottomrule
\end{tabular}
}
\end{table}

\begin{figure}
    \centering
    \includegraphics[width=.8\textwidth]{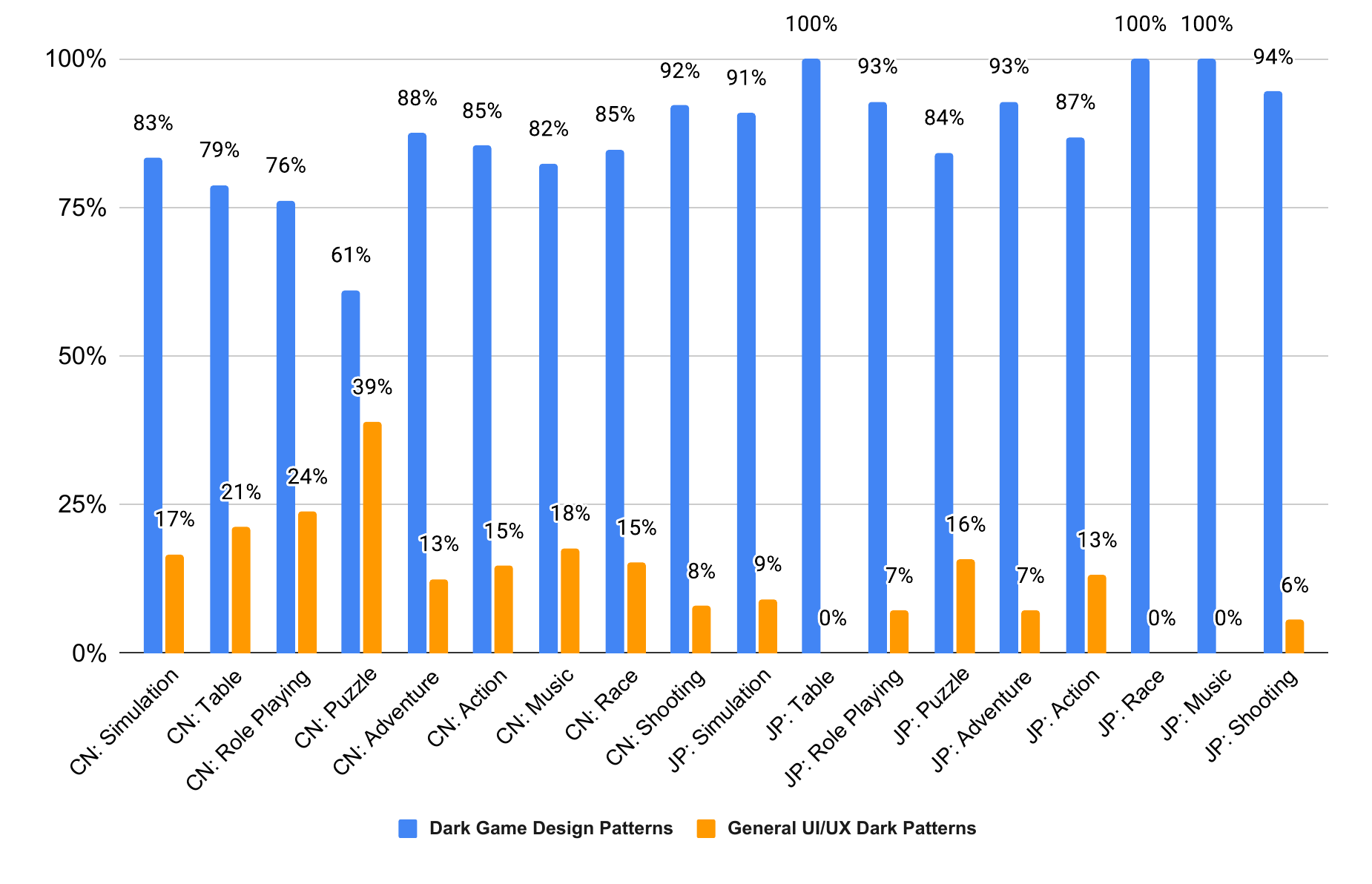}
    \caption{Percentage comparison of game DPs vs. general UI/UX DPs. CN: Chinese. JP: Japanese.}
    \label{fig:gamedpvsgdp}
    \Description{A figure showing the percentages for game-based and generic DPs. Overall, there were more game-based patterns, but the Chinese puzzle game featured closer counts (61\% game-based and 39\% generic).}
\end{figure}

We found that most patterns were game DPs 
(\autoref{fig:gamedpvsgdp}). The general UI/UX DPs we identified included \textbf{Bad Defaults}, \textbf{False Hierarchy}, \textbf{Visual Prominence}, \textbf{Alphabet Soup}, \textbf{Attention Capture}, \textbf{Privacy Zuckering}, \textbf{Parasocial Pressure}, \textbf{Sneaking}, and \textbf{Partitioned Pricing}. These patterns were largely independent of the game mechanics, i.e., non-game-specific UI/UX and interaction elements, which aligns with \citet{gray2024ontology}'s findings. Since DPs are context-dependent, we primarily focused on game-specific varieties. Designers have the creative freedom to construct immersive game worlds. Game DPs thus reflect the dynamics of the game environments and narratives within which they are embedded. All previously identified high-level game DPs 
were found. We identified a new high-level pattern, \textbf{Technical Deceptive Game Design Pattern}, along with a new meso-level pattern, \textbf{Fragmented Downloads}. We also discovered several new low-level patterns, including \textbf{Polymorphic Currency}, \textbf{Remedy Consumption}, \textbf{Risk Conversion}, \textbf{Bundle Bonus}, \textbf{First Charge Discount}, \textbf{Accumulating Rewards}, \textbf{Overloading}, and \textbf{Sycophant}. We provide definitions and detailed examples next.

\subsubsection{Monetary Patterns}
We identified several new patterns, including a new meso-level pattern called ``Never-Ending Lure'' that has two low-level patterns: ``First Charge Discount'' and ``Accumulating Rewards.'' We also identified ``Risk Conversion'' and ``Bundle Bonus'' as new low-level patterns within the ``Gambling'' category, as they emerge in conjunction with gambling mechanisms in games. Similarly, ``Remedy Consumption,'' a new low-level pattern under the ``Deceptive Luxury'' category, often occurs in combination with ``Artificial Scarcity.'' Finally, we identified ``Polymorphic Currency'' as a new low-level pattern within the ``Intermediate Currency'' category.

\textbf{``First Charge Discount''} happens when in-game items or currencies are offered at significantly reduced prices for the player's first purchase, and then revert to the original higher price for subsequent purchases. This pattern distorts the player's perception of in-game items' true value. Unlike \textbf{Premium Currency}, it focuses on a one-time-only purchase to mislead players into feeling that they have taken advantage of the purchase.

\textbf{``Accumulating Rewards''} involves granting privileges, special offers, and rewards to players who spend above certain monetary thresholds. Unlike \textbf{Pay to Win}, where players directly spend money to gain in-game advantages, this pattern does not emphasize helping players progress faster. Instead, it offers players a special, privileged, and extraordinary status that is visible to other players. For instance, during the recording session, we encountered a game that encouraged players to spend money, accumulating towards a certain VIP level. Instead of directly paying for higher VIP levels, this system tracks all in-game spending behaviours---including purchases of premium currency, battle passes and other items---gradually accumulating their value to determine the player's VIP level. Different VIP levels unlocked different powerful characters and unique aesthetic items that could only be accessed through continued spending.

\textbf{``Risk Conversion''} offers players an alternative way to achieve their goals, such as obtaining an extremely rare item, by spending a substantial amount of money to circumvent the uncertainties inherent in gambling mechanics. However, the actual amount players spend is significantly higher than if they had opted for gambling mechanics. This pattern specifically targets risk-averse players.

\textbf{``Bundle Bonus''} encourages players to engage in multiple gambling attempts at once rather than a single one. For instance, if a player undertakes 10 in-game lottery activities at once, they receive an additional opportunity for free, incentivizing bulk gambling.

\textbf{``Remedy Consumption''} is a pattern where players are given a second opportunity to correct a missed purchase of some limited, valuable in-game items. This involves intentional artificial scarcity for these in-game items, which compels players to expend additional effort or incur further costs (compared to the first time) to obtain them.

\begin{figure}[!ht]
    \centering
    \includegraphics[width=\textwidth]{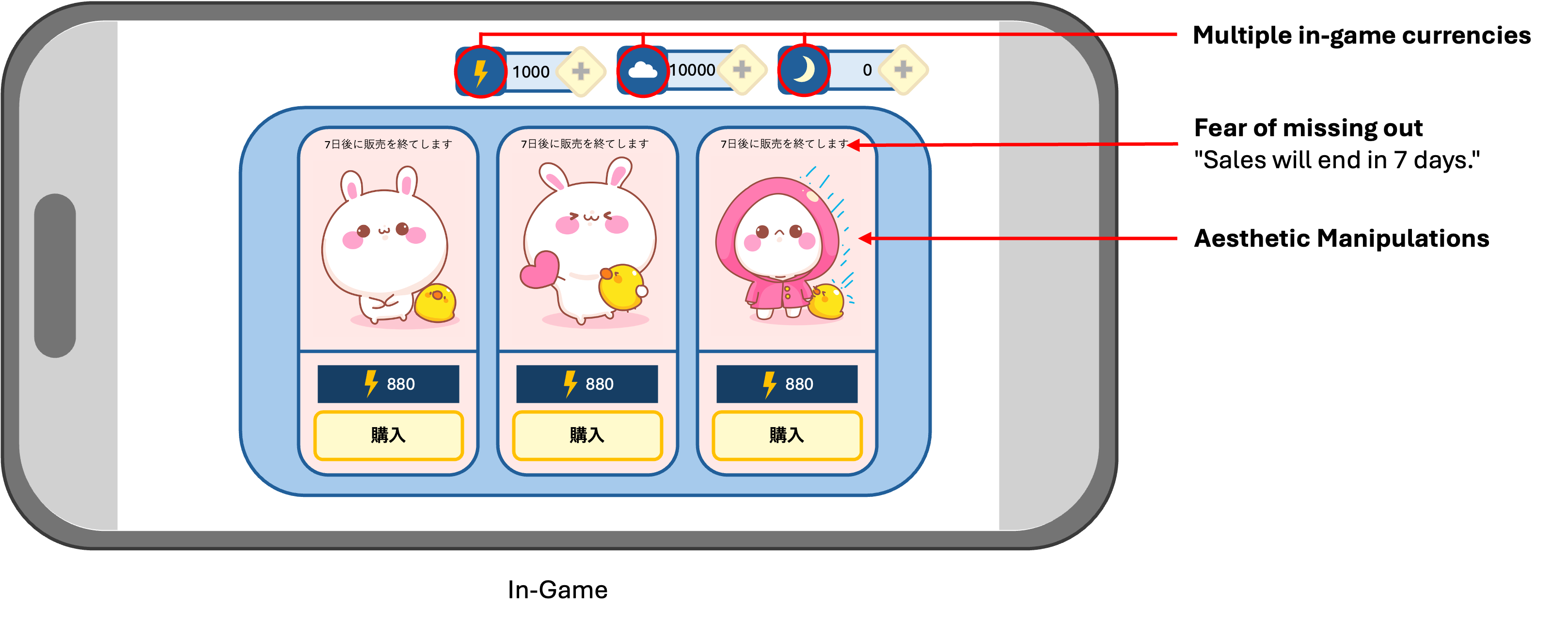}
    \caption{\textbf{``Polymorphic Currencies''} is the use of multiple, purpose-specific in-game currencies that must be purchased with real-world money and are non-interchangeable, often resulting in leftover amounts.
    \href{https://www.freepik.com/free-vector/set-cute-rabbit-with-duck-different-posture-cartoon-illustration_12573652.htm\#fromView=search&page=2&position=0&uuid=7b633344-548b-4b44-9417-99cef20a799e}{Rabbit Image © mamewmy, Freepik}.}
    \Description{Illustration of the Polymorphic Currencies pattern, which shows multiple currencies in the game menu. Fear of Missing Out (FOMO) and Aesthetic Manipulations are also present in the use of sad-looking characters beside time-limited options.}
    \label{fig:polymorphiccurrencies}
\end{figure}

\textbf{``Polymorphic Currency''} is the use of multiple, purpose-specific in-game currencies within a single game that can either be purchased with real-world money or earned by in-game activities. The malicious part is that they are non-interchangeable, unlike \textbf{Premium Currency}, designed to obscure the exchange rate between real-world money and in-game currencies. This 
burdens the player 
and leads them to eventually spend more. \autoref{fig:polymorphiccurrencies} illustrates an example. Here, ``Lightning Coins'' are designated for purchasing character skins, ``Gold Coins'' are used to exchange for common in-game items, and ``Moon Icons'' serve as a special currency for acquiring event-exclusive items, expiring when the event concludes. Importantly, all three currencies can be obtained through real-world money transactions; however, they are not interchangeable within the game, players have to spend separately to get different kinds of coins, which leads to the frequent accumulation of unused and leftover currencies.

\subsubsection{Psychological Patterns}
We introduced the meso-level pattern ``Reward Mania,'' which encompasses two newly discovered low-level patterns: ``Overloading'' and ``Sycophant,'' as well as the previously recognized ``Variable Rewards''~\cite{sousa2023dark}. 

\begin{figure}
    \centering
    \includegraphics[width=1\textwidth]{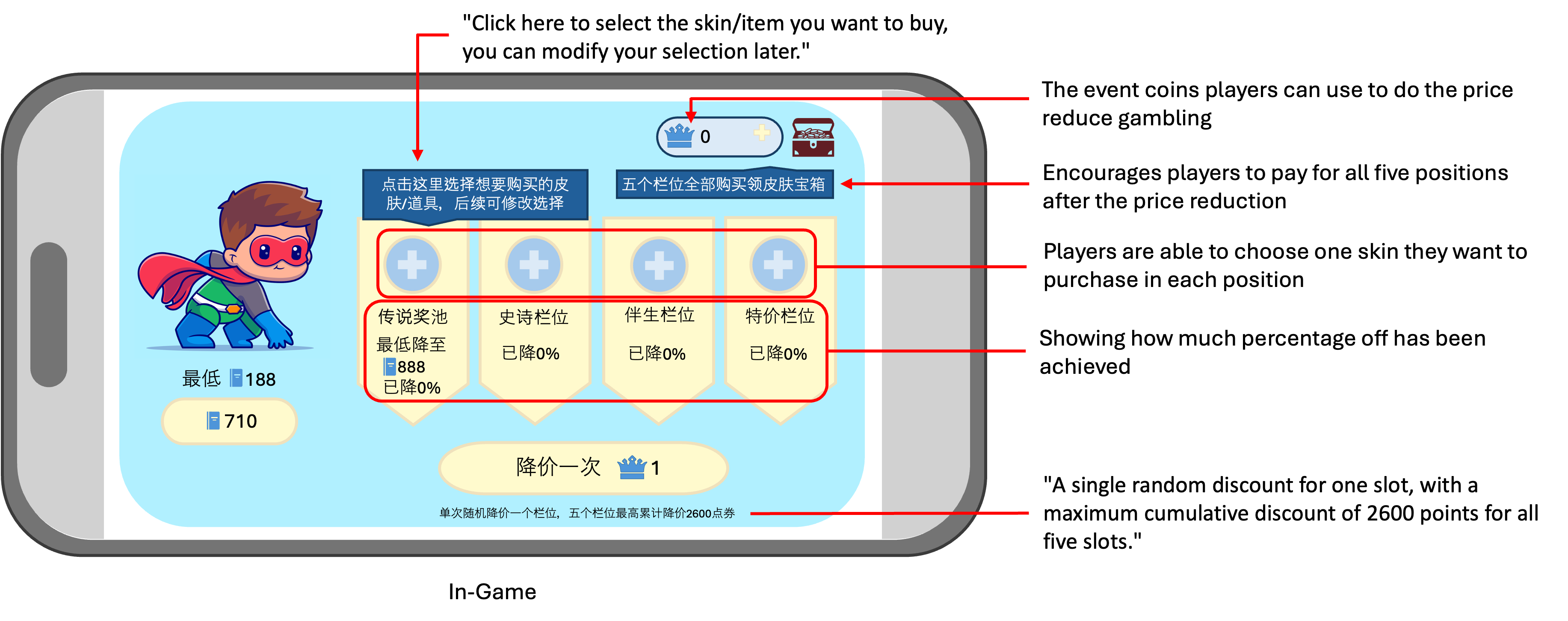}
    \caption{\textbf{``Overloading''} pattern occurs when in-game reward mechanisms are designed to be overly complex, enticing the player to invest effort in understanding them. 
    \href{https://www.freepik.com/free-vector/cute-man-super-hero-landing-cartoon-vector-icon-illustration-people-holiday-icon-isolated flat_230902273.htm\#fromView=search&page=1&position=3&uuid=ae47daf8-bced-4fe6-b008-c46520527e80}{Hero image © catalyststuff, Freepik}.}
    \Description{Illustration of the Overloading pattern, which shows the complexity of an avatar skin selection mechanism in the game menu. For example, random discounts are offered for one of five slots but have a maximum cumulative effect.}
    \label{fig:overloading}
\end{figure}

\textbf{``Overloading''} occurs in in-game mechanisms, such as reward systems or gambling, that are designed to be overly complex, thereby enticing players to invest significant effort in understanding and engaging with them. The example in  \autoref{fig:overloading} illustrates this pattern, where players are enticed to purchase in-game character skins at a discounted price. The interface features five options, one set to a default skin; players can select four additional skins under certain constraints. In many games, cosmetic modifications, such as character skins, are categorized by rarity levels~\cite{luo2023optimal}. In this instance, the skins are arranged with the highest rarity level on the far left and the lowest on the far right. Players must collect ``crown coins'' to unlock discounts. Each crown coin applies a random discount to one of the five options, but there is a maximum discount. Crown coins can be earned by completing daily tasks or by direct purchase. Purchasing all five skins is incentivized, as it grants an extra skin loot. This process demands significant cognitive effort from players, particularly if they aim to maximize the discount on the highest rarity skin while securing the best overall discount.

\begin{figure}
    \centering
    \includegraphics[width=\textwidth]{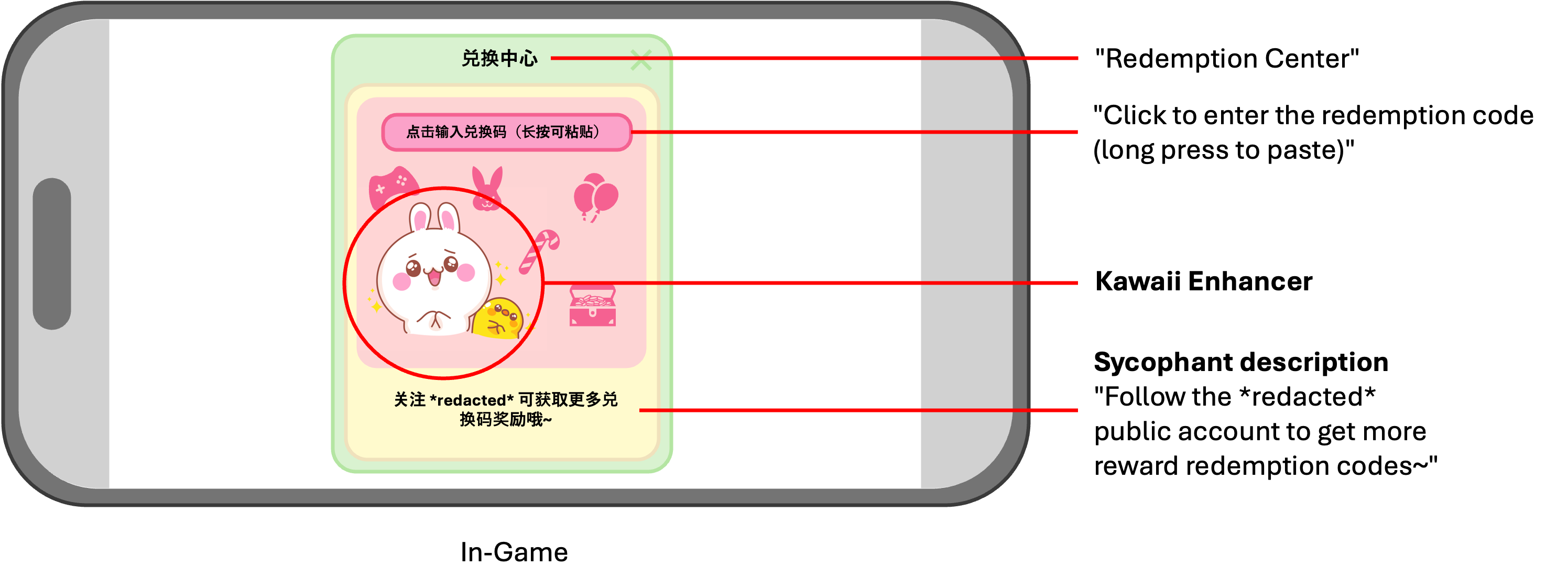}
    \caption{The \textbf{Sycophant} pattern uses in-game rewards to incentivize out-of-game social activities, such as following the game's social media accounts or participating in related online communities. 
    \href{https://www.freepik.com/free-vector/set-cute-rabbit-with-duck-feel-happy-sad-cartoon-illustration_12573654.htm\#query=kawaii\%20character&position=12&from_view=keyword&track=ais_hybrid&uuid=40c03a88-3d03-4516-8e49-612c7158d1ad}{Rabbit image © mamewmy, Freepik}.}
    \Description{Illustration of the Sycophant pattern with a kawaii enhancer, which shows a cute character enticing the player to follow their SNS account.}
    \label{fig:sycophantkawaii}
\end{figure}

\textbf{``Sycophant''} refers to the use of rewards within the game to encourage players to allocate additional attention and resources to social activities outside of the game, such as following the game's social media accounts or participating in related online communities.  \autoref{fig:sycophantkawaii} illustrates an example of the ``Sycophant'' pattern. In this scenario, players are encouraged to follow the game developers' or the game's social media accounts in the hope of obtaining a redemption code, which can be exchanged for in-game rewards. This example also incorporates additional mechanisms that further reinforce the DP, which will be discussed in subsequent sections.

\subsubsection{Technical Patterns (New)}
We identified a new high-level pattern, the \textbf{Technical} pattern, which deceives users or players by obscuring or misrepresenting the technical aspects of the game application. Within this category, we discovered a single subcategory, ``Fragmented Downloads.''

\begin{figure}
    \centering
    \includegraphics[width=.86\textwidth]{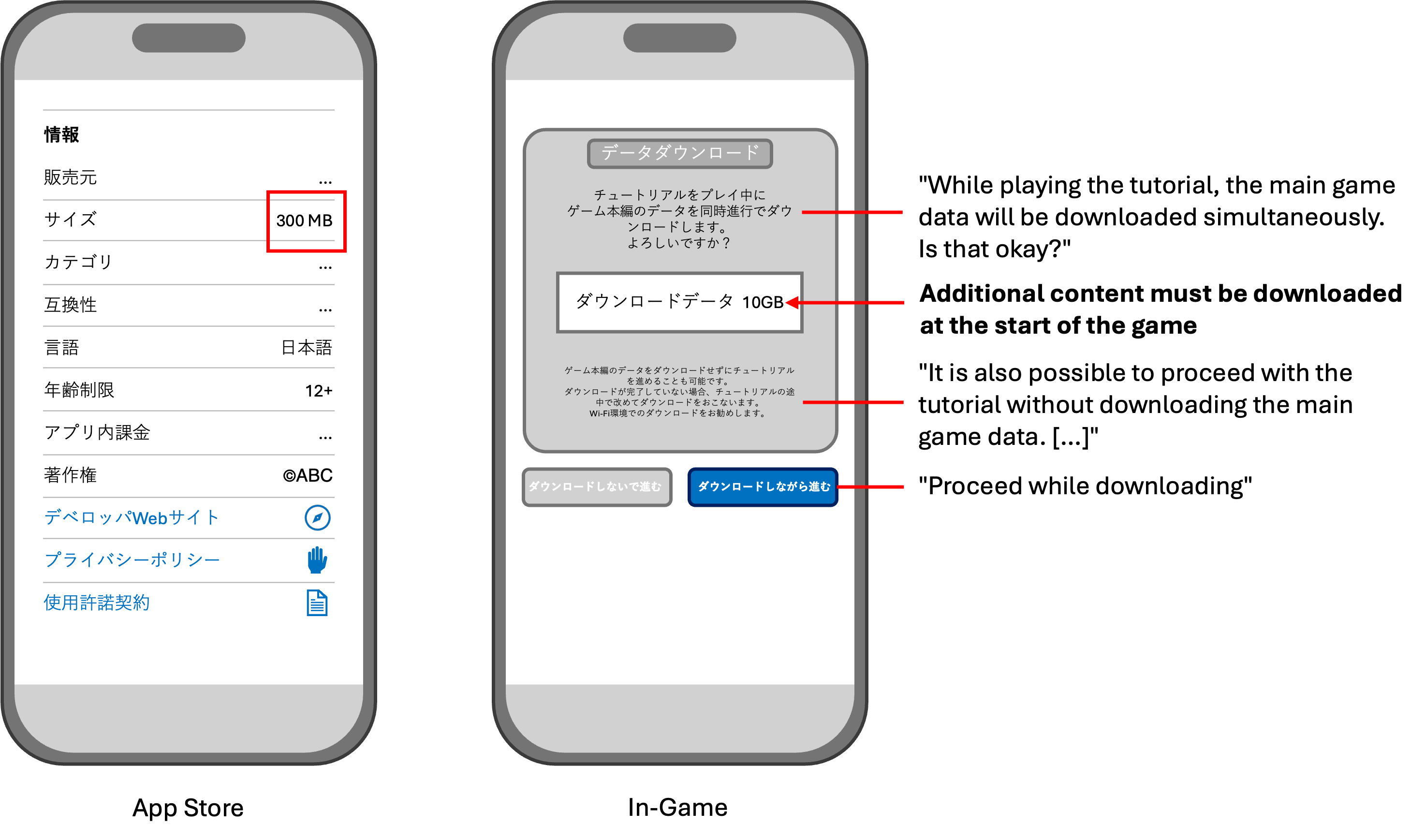}
    \caption{The \textbf{Fragmented Downloads} pattern conceals the actual storage space needed for playing until the game begins.}
    \Description{Illustration of the Fragmented Downloads pattern. Two screens are shown to compare the download size of the game in the app store and the additional downloads needed that are only presented in-game.}
    \label{fig:fragmenteddownloads}
\end{figure}

\begin{figure}
    \centering
    \includegraphics[width=\textwidth]{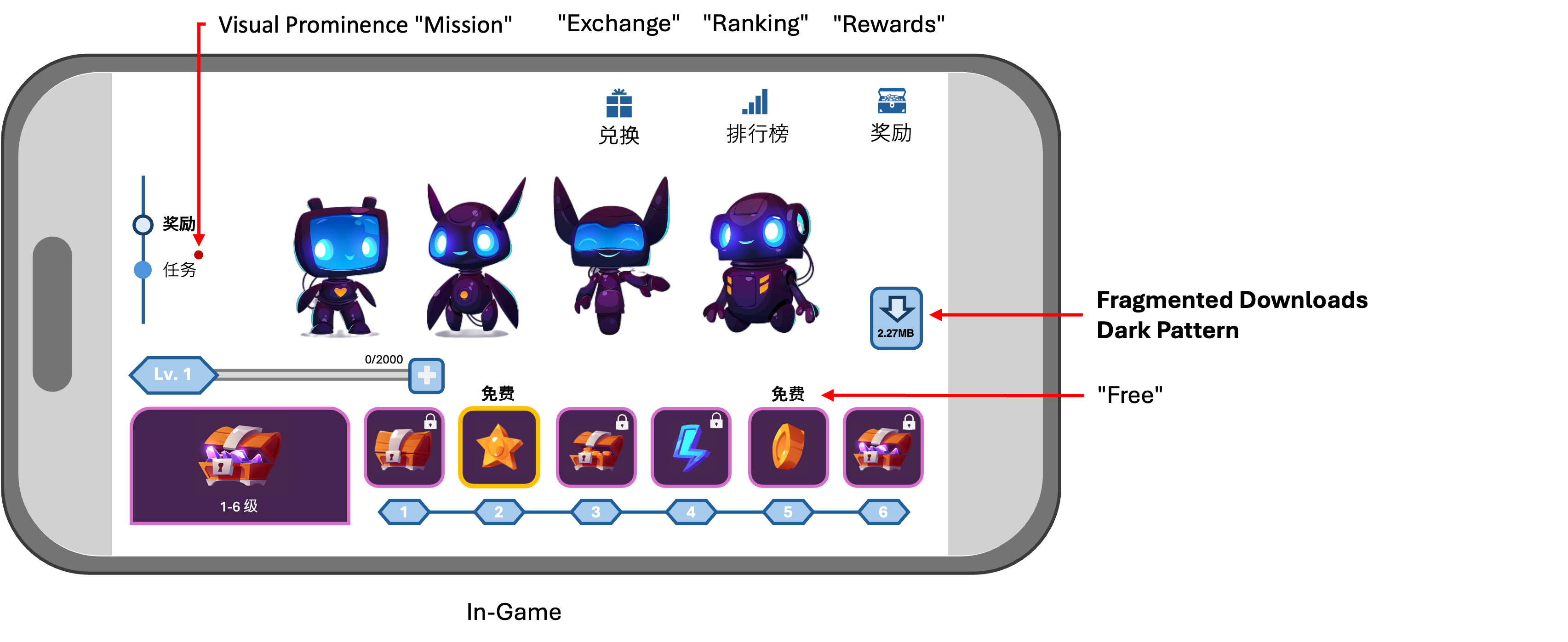}
    \caption{The \textbf{\emph{Evil} Battle Pass} pattern occurs with misuse or overuse of battle passes. Players are required to spend excessive time completing tasks, risking expiration of the battle pass and loss of rewards, even if they have paid for it in advance. 
    \href{https://www.freepik.com/free-vector/game-icon-kit-cartoon-vector-gui-assets-set-lock-key-blue-lightning-first-aid-medicine-bag-various-gem-stones-closed-open-wooden-chest-box-with-treasure-hp-heart-time-trophy_73605686.htm\#fromView=search&page=2&position=34&uuid=641a15d0-97db-400f-ae45-0c9337310c28}{Battle Pass Items image} and
    \href{https://www.freepik.com/free-vector/set-robot-mascots-isolated-background_154651791.htm\#fromView=search&page=2&position=30&uuid=dd09eeda-0e06-446f-9c64-c817fc6ca6a2}{Robot Character image © upklyak, Freepik}.}
    \Description{Illustration of the \emph{Evil} Battle Pass, which shows a battle pass with ``free'' items that require extensive time and effort for access to be maintained.}
    \label{fig:evilbattlepass1}
\end{figure}

\textbf{``Fragmented Downloads''} are characterized by designs that obscure the true storage requirements of the game until the game starts. This is deceptive because it targets mobile users who are cautious about their device's storage capacity and foments a space monopoly by the app against competitors. Users might initially avoid downloading a mobile game that requires excessive space. We encountered two distinct scenarios. The first scenario, illustrated in  \autoref{fig:fragmenteddownloads}, allows players to choose whether to download the main content while playing the tutorial session. Although declining is possible, the download prompt repeatedly appears during the tutorial. If they do not download the content, they are unable to progress to the main portion of the game. Another scenario involves the game omitting elements that typically require more memory, such as animations. As in \autoref{fig:evilbattlepass1}, if players choose not to download the additional packages, certain animations---like those for character selection before a match or viewing the details of skins (cosmetic modifications)---will not appear on the screen. Instead, a ``downloading'' button will prompt players to download the missing content. Although players can continue gameplay without downloading these contents, their experience may feel incomplete due to the missing visual elements on the screen.

\begin{figure}
    \centering
    \includegraphics[width=\textwidth]{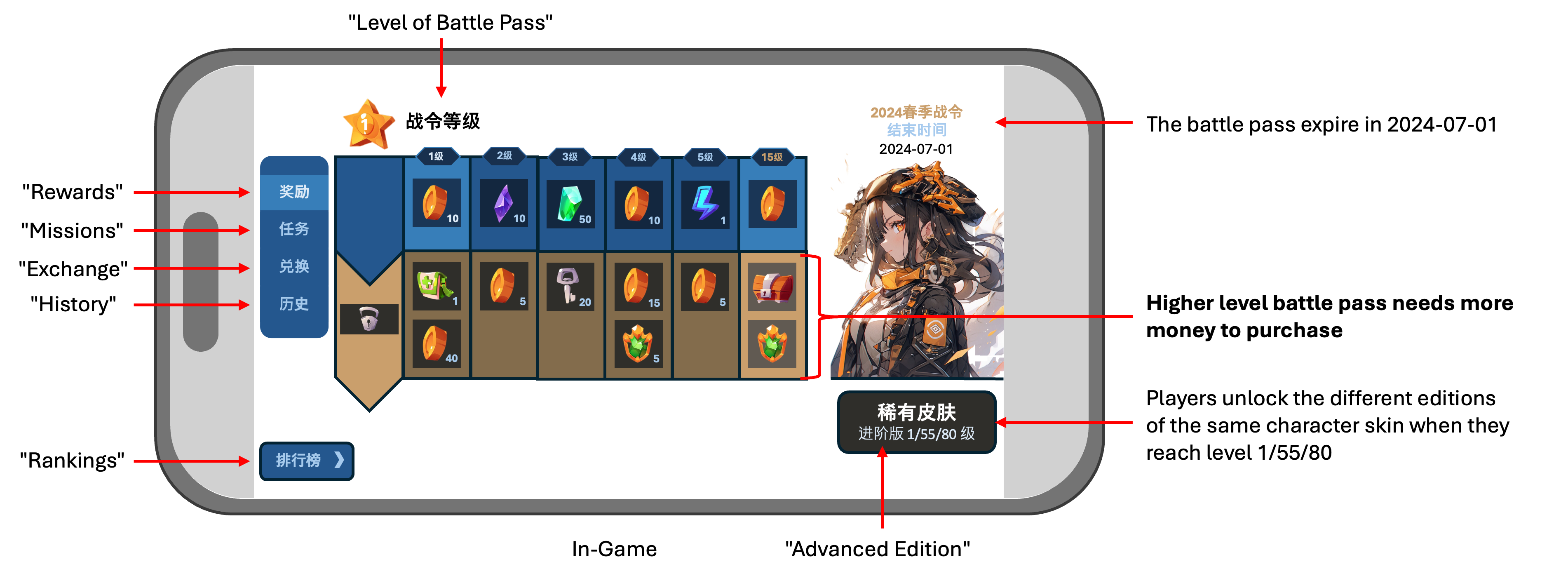}
    \caption{Misuse or overuse of battle passes occurs when players are required to spend excessive time completing tasks, risking expiration of the battle pass and loss of rewards, even if they have paid for it in advance. 
    \href{https://www.freepik.com/free-vector/game-icon-kit-cartoon-vector-gui-assets-set-lock-key-blue-lightning-first-aid-medicine-bag-various-gem-stones-closed-open-wooden-chest-box-with-treasure-hp-heart-time-trophy_73605686.htm\#fromView=search&page=2&position=34&uuid=641a15d0-97db-400f-ae45-0c9337310c28}{Battle Items image © upklyak, Freepik};
    \href{https://www.freepik.com/free-ai-image/anime-character-with-dragon-illustration_114338568.htm}{Battle Pass Character image © Freepik}.}
    \Description{Illustration of the \emph{Evil} Battle Pass pattern, which shows an additional fee-based battle pass.}
    \label{fig:evilbattlepass2}
\end{figure}

\subsection{Meta-Level Designs that Augment Deception 
}
We identified two meta-level design strategies that influenced the effectiveness of DPs. 

\subsubsection{DP Combos}
Some low-level patterns overlapped with other DPs. For example, ``\emph{Evil} Battle Pass'' is categorized under the ``Playing by Appointment'' meso-level pattern, but its design also incorporates elements of ``Grinding'' and ``Daily Rewards,'' encouraging players to log in and complete daily tasks to earn experience points (XP) and level up their battle passes. Similarly, rather than relying on a single DP to deceive or manipulate players for a single purpose, game developers sometimes employed multiple DPs simultaneously within a single design or interface towards a particular end. We refer to these as \textbf{DP Combos}. For instance, when we first launched a certain game, we were presented with the choice to log in as a guest or through their social media accounts, with a note that guest logins come with limitations and a risk of data loss. If we chose to log in via social media, we encountered a combination of ``Bad Defaults'' and ``Friend Spam/Impersonation,'' along with a ``False Hierarchy,'' where the critical ``Terms of Use Agreement'' was hidden among other less important marketing or notification permissions. Another example was the combined use of ``Premium Currency'' and ``Bulk Purchase.'' Instead of allowing players to purchase an exact amount of intermediate currency, they are forced to choose from predetermined bulk amounts. The higher the bulk amount, the better the exchange rate between virtual currency and real-world money, thereby manipulating players into spending more.

As we were exploring game DPs in onboarding phase, we identified \textbf{``Trapping Starter Kits''} Combo that combined the patterns ``Playing by Appointment,'' ``Daily Rewards'' and ``Grinding.'' This DP Combo particularly entices new players to log in and play continuously for the first few days after creating new accounts. These designs often use in-game rewards as incentives, creating a situation where new players lose the opportunity to obtain these rewards if they fail to log in and complete tasks within the specified timeframe. Another DP Combo we frequently observed was  \textbf{``Try Now and Pay Later,''} this Combo combined ``Grinding,'' ``Pay to Win,'' ``Pay Wall,'' and ``Easy to Get, Hard to Lose.'' This DP Combo offers players the chance to initially try premium experiences, powerful items, or privileges, with payment required at a later stage. For example, a game might provide a trial experience with powerful characters, allowing players to complete tasks easily and quickly. However, if players wish to permanently acquire these characters, they must either spend money or complete numerous tasks, requiring significant additional time.

\subsubsection{DP Enhancers}
Many games incorporate elements of cuteness-as-kawaii in the design of virtual items, characters, or aesthetics, much like in real life. While a neutral concept, cuteness has been identified as a low-level pattern under the meso-level category of ``Emotional or Sensory Manipulation''~\cite{gray2024ontology, luguri2021shining, lacey2019cuteness}. However, when examining our sample, we observed that most such designs were not instances of DPs, but deployed alongside \emph{other} DPs like ``Aesthetic Manipulation'' to enhance their effects. We call this use of kawaii a \textbf{DP Enhancer}: designs that are not inherently deceptive or manipulative but can intensify the effects of accompanying DPs. In our corpora, we identified several DP Enhancers, including ``Cuteness/Kawaii,'' but also ``The Power of NPCs,'' ``Rarity Level,'' and ``Functional Value.''

``\textbf{The Power of NPCs}'' occurs when neutral non-player characters (NPCs) are deployed for deceptive purposes. For instance, we encountered a tutorial session led by a popular NPC, who guided us on how to play, explained interface functions, and demonstrated how to engage in gambling mechanisms, like obtaining powerful characters through a lottery system, with a free trial. This interaction significantly enhanced ``Optimism and Frequency Biases,'' leading us to overestimate the likelihood of certain events or occurrences due to the recent or memorable exposure.

``\textbf{Cuteness/Kawaii}'' (\autoref{fig:sycophantkawaii}) acts as a DP Enhancer when used in conjunction with the ``Sycophant'' low-level \textbf{Psychological} pattern. 
Kawaii characters were employed to persuade players to follow the game's official social media account for additional rewards.

``\textbf{Rarity Level}'' reflects how in-game items are often designed with varying levels of rarity. For instance, in Chinese simulation games, character skins are ranked from B rank to S++ rank, with higher ranks featuring more intricate designs and commanding higher prices, as we observed. Rarity levels were often combined with the ``Deceptive Luxury'' meso-level \textbf{Monetary} pattern, where the higher the rarity, the scarcer and more desirable the item becomes.

``\textbf{Functional Value}:'' Players rely on functional values such as attack (ATK), defence (DEF), and health (HP) to determine their ability to overcome dungeons, monsters, and other players, making these statistical advantages crucial. We identified the use of functional value as a means to enhance the ``Complete the Collection'' meso-level \textbf{Psychological} pattern. Specifically, the more characters players collect---often through direct purchases or gambling mechanisms like loot boxes---the greater the statistical bonuses they receive, such as increased ATK and HP.

\subsection{Distribution of 
DPs (RQ2)}

We identified an average of 21.11 game DPs (SD: 9.65) and 3.39 general UI/UX DPs (SD: 2.62) per game. \textbf{Monetary} patterns 
were the most prevalent (M=8.06, SD: 3.92). Most (54 out of 380 game DPs) were associated with ``Intermediate Currency,'' where players use real-world money to acquire in-game currency for purchases or trades within the game world. Specifically, 15 out of 54 Monetary DPs were linked to ``Premium Currency.'' 
Additionally, 20 out of 54 Monetary DPs were related to the ``Polymorphic Currency,'' a newly identified pattern, while 19 out of 54 Monetary DPs are tied to ``Bulk Purchase,'' where players can only buy in-game currencies in fixed bulk amounts~\cite{king2023investigating}.

\textbf{Psychological} patterns 
were the second most prevalent (average of 5.78 patterns per game, SD: 3.47). ``Aesthetic Manipulation,'' which involves using misleading questions or exploiting emotional and subconscious desires to influence player behaviour, accounted for 27 out of 104 Psychological DPs. The ``Complete the Collection'' pattern, which drives players to feel compelled to acquire all in-game items, achievements, or myths (secret achievements for the player to discover), made up 25 out of 104 Psychological DPs. The ``Easy to Get, Hard to Lose'' meso-level pattern accounted for 18 out of 104 Psychological DPs, including eleven ``Invested/Endowed Value,'' two ``Endowed Progress,'' 
and five ``Try Now and Pay Later'' combo patterns. A newly identified meso-level pattern, ``Reward Mania,'' accounted for 26 out of 104 Psychological DPs, comprising six ``Overloading,'' fifteen ``Sycophant,'' and five ``Variable Rewards.'' 

\textbf{Temporal} patterns 
were common (M=5.06, SD: 2.73). The ``Playing by Appointment'' meso-level pattern was the most dominant, accounting for 48 out of 91 Temporal DPs, including nineteen attributed to ``Others,'' twenty ``\emph{Evil} Battle Pass,'' none to ``Wait to Play'' low-level patterns, and nine ``Trapping Starter Kit'' combo patterns. ``Daily Rewards'' constituted 19 out of 91 Temporal patterns, manifesting primarily as daily log-ins and daily tasks.

An average of 1.61 \textbf{Social} patterns were identified per game (SD: 1.24). 
10 out of 29 were ``Social Pyramid Scheme'' low-level patterns, where players were incentivized to invite friends to the game, who were then also encouraged to invite others to earn rewards~\cite{zagal2013dark}. ``Friend Spam/Impersonation'' meso-level patterns accounted for 10 out of 29 Social DPs, involving unsolicited notifications and messages sent by the game to players' social media accounts and contact lists~\cite{zagal2013dark}. Another 7 out of 29 Social DPs were attributed to ``Fear of Missing Out (FOMO)'' meso-level patterns, where the game deliberately creates a sense of urgency by implying that pausing or quitting the game would lead to missed rewards or an inability to keep up with other players~\cite{sousa2023dark}, with two Social DPs were attributed to ``Competition.'' 

We also identified a new high-level pattern, \textbf{Technical} patterns, which includes the ``Fragmented Downloads'' meso-level pattern as a subclass. This pattern was observed in half of 18 games. The pattern either creates a ``download wall'' that stops users from continuing if they do not download the necessary components or shows an incomplete visual experience if specific resource packages are not downloaded. 

\subsection{Comparison of the Two Corpora: China and Japan (RQ3)}
We identified general UI/UX DPs across the corpora. One common pattern was ``Visual Prominence,'' which involves displaying visually prominent symbols, such as red dots, within certain UI---like mission screens and in-game stores---to grab player attention in pursuit of checking for unfinished tasks or viewing newly launched items for purchase. We also identified ``Parasocial Pressure.'' Since the selected games were not single-player, players often had opportunities to observe the behaviour or achievements of others, whether through direct competition or by reading announcements that congratulated other players' successes. For example, players can send instant in-game posters to their friends to showcase their recently purchased skins. When recipients click on these posters, they are immediately directed to the item's shop page. Additionally, patterns such as ``Privacy Zuckering,'' ``Sneaking,'' and ``Partitioned Pricing'' were found \emph{exclusively} in the Chinese corpus. ``Partitioned Pricing'' was identified when switching the language setting from Chinese to English, resulting in variations in the discount offered for purchasing premium currency. 

The Chinese corpus showed an average of 5.78 \textbf{Temporal} patterns per game (SD: 2.82). In comparison, the Japanese corpus had an average of 4.33 \textbf{Temporal} patterns per game (SD: 2.60). The ``Playing by Appointment'' meso-level pattern was dominant in both corpora. 
A notable feature of the Chinese corpus was the frequent appearance of multiple ``\emph{Evil} Battle Pass'' patterns within the same game, indicating reliance by some developers on this pattern. \autoref{fig:evilbattlepass1} and \autoref{fig:evilbattlepass2} illustrate two examples found within the same game.
Both include a limited number of free, low-value rewards, but players must purchase each to access the full collection of rewards, which are more valuable and rarer. ``Trapping Starter Kit,'' ``Daily Rewards,'' and ``Advertisement'' patterns were identified in both corpora, while patterns such as ``Wait to Play,'' ``Infinite Treadmill,'' and ``Can't Pause or Save'' were all \emph{absent}. 

When it came to \textbf{Monetary} patterns, the Chinese corpus had more instances than the Japanese corpus. In the Chinese corpus, there were an average of 9.78 per game (SD: 3.67), while in the Japanese corpus, there were an average of 6.33 per game (SD: 3.54). The ``Intermediate Currency'' meso-level pattern was the most prevalent across both corpora, with all games employing ``Premium Currency'' and ``Bulk Purchase'' patterns. ``Polymorphic Currency'' was found in all games except for puzzle games. Notably, we could not search for the ``Leftovers'' pattern, since we did not engage in purchasing or spending intermediate currencies during our analysis. Another common pattern was the ``Gambling'' meso-level pattern, which includes ``Loot Boxes,'' ``Risk Conversion,'' and ``Bundle Bonus,'' used in both Chinese and Japanese corpora to encourage players to spend money on highly powerful, valuable, and rare items. Additionally, ``Pay to Skip'' and ``Anchoring Tricks'' patterns were identified in both corpora. Notably, the ``Deceptive Luxury'' meso-level pattern, 
and the ``Never-Ending Lure'' meso-level pattern 
were found only in the Chinese corpus.

On average, 1.89 \textbf{Social} patterns were identified in the Chinese corpus (SD: 1.54), with 1.33 patterns in the Japanese (SD: 0.87) corpus. ``Social Pyramid Scheme'' low-level patterns were found in both corpora, while ``Friend Spam/Impersonation'' meso-level patterns only appeared in the Chinese corpus. Although there is some overlap between ``\emph{Evil} Battle Pass'' and ``FOMO'' patterns---battle passes are seasonal events with exclusive rewards like seasonal skins---we did not find other FOMO patterns in the Chinese corpus. In contrast, the FOMO low-level pattern was common in the Japanese corpus.

The Chinese corpus had a higher average of 7.44 \textbf{Psychological} patterns per game (SD: 4.13), with the Japanese corpus having an average of 4.11 (SD: 1.54). ``Aesthetic Manipulation'' emerged as the predominant pattern in both corpora. Also, the patterns ``Complete the Collection,'' ``Invested/Endowed Value,'' ``Optimism and Frequency Biases'' were observed across corpora. However, certain patterns, such as the low-level patterns like ``Overloading'' and ``Variable Rewards,'' along with ``Illusion of Control'' and ``Endowed Progress,'' were unique to the Chinese corpus.

For the newly identified \textbf{Technical} high-level pattern, our analysis of ``Fragmented Downloads'' revealed its common presence in both Chinese and Japanese corpora.

\section{Discussion}
\label{sec:discussion}
Our investigation of the onboarding experience provided by free-to-play mobile games in the Chinese and Japanese app markets 
revealed several important findings for game-based DPs. We confirmed the presence of known high-level game DPs 
and determined how certain newly found DPs manipulate players in uncharted ways (RQ1). We mapped these distributions across corpora (RQ2 \& RQ3).
We also discovered that game developers often employ a combination of DPs in practice, and that DP Enhancers 
can amplify the effects of DPs. We now discuss the findings
.

\subsection{An Enriched Ontology for Deceptive Game Design Patterns}

For years, researchers have been studying the malicious, deceptive, and manipulative UI/UX patterns across various domains~\cite{karagoel2021dark, mathur2019dark, zagal2013dark,gray2024ontology}. 
Following previous research~\cite{sousa2023dark, dahlan2022finding, king2023investigating, hadan2024}, 
we drew on but also extended the game-oriented framework by \citet{zagal2013dark} by bringing in other game frameworks~\cite{sousa2023dark} and lay resources\footnote{\url{https://www.darkpattern.games}}, as well as leveraging the general taxonomy by \citet{gray2024ontology}.
By exploring the selected sample games,  
we were able to find new game DPs. 

The game DPs we found often involved multiple aspects of deception or manipulation beyond the unitary categories proposed by \citet{zagal2013dark}. This is consistent with recent studies~\cite{hadan2024, king2023investigating, sousa2023dark}. \citet{hadan2024} 
revealed that the Battle Pass system incorporates various manipulative elements such as ``Grinding,'' ``Infinite Treadmill,'' and ``Anchoring Tricks.'' 
Yet, existing frameworks did not adequately address the multifaceted nature of deception surrounding a single DP or multiple DPs. 
In considering the primary intention behind battle passes, which is to ensure players remain persistently engaged and attentive to the game~\cite{joseph2021battle}, we classified ``\emph{Evil} Battle Pass'' under the meso-level ``Playing by Appointment.'' 
We applied a similar rationale to incorporate all DPs identified in previous research~\cite{zagal2013dark, sousa2023dark, king2023investigating}, public resources such as the Dark Pattern Games website\footnote{\url{https://www.darkpattern.games}}, and newly discovered game DPs into the current three-level ontology (\autoref{table:fradp}). The following subsections offer detailed case study discussions on the new DPs.

\subsection{New Classes and Subclasses 
of Deceptive Game Design}
We introduced a new high-level pattern, the \textbf{Technical} pattern, along with several meso-level patterns, including ``Never-Ending Lure'' 
and ``Fragmented Downloads.'' At the low-level, we identified specific patterns such as 
``Polymorphic Currency,'' ``Remedy Consumption,'' ``Risk Conversion,'' ``Bundle Bonus,'' ``First Charge Discount,'' ``Accumulating Rewards,'' 
``Overloading,'' and ``Sycophant.'' 

Mobile games possess technical features that developers can exploit to deceive, manipulate, and trick users. We identified the meso-level pattern ``Fragmented Downloads,'' where users are deceived about the true storage requirements of the game before the gameplay begins. 
Companies face pressure to meet the demand of marketing to new segmented audiences, which necessitates novel strategies~\cite{politowski2021game}. 
People also use mobile devices for multiple purposes~\cite{jin2008mobile}. 
Companies may exploit low user awareness of app space usage and requirements by promoting downloads that contribute to the game's ranking metrics. 
Ultimately, this can lead to the exploitation of user wealth, and should be challenged.

People often expect that games offering ``Free-to-Play'' and ``microtransaction'' setups will remain free~\cite{fitton2019creating}.
While microtransactions often serve as equivalent to traditional cheating~\cite{tomic2017effects,freeman2022ingamepurch}, 
combo patterns like ``Trapping Starter Kits'' and ``Try Now and Pay Later'' deceive players into believing that they can achieve similar results without payment, while instead manipulating them into investing excessive effort. 
This highlights the trade-off between time and money that players face in games~\cite{petrovskaya2021predatory, Petrovskaya2022micro}. As \citet{yee2006labor} observed, such tactics increasingly blur the line between play and work, conditioning players to become more efficient ``game workers'' by demanding greater effort. 

Reward elements have been linked to problematic gameplay behaviours, 
because they make players voluntarily invest in bold and persistent efforts~\cite{pirrone2024we, butler2015applied}. 
The newly identified pattern ``Overloading'' occurs when game mechanics exploit the player's cognitive capacities. \citet{xiao2022probability} demonstrated that excessively complex decision-making environments may cause players to account for all relevant information, leading to poor choices. While \citet{mathur2021what} confirmed that DPs can impose cognitive burdens, 
our analysis further revealed that achieving optimal outcomes is nearly impossible due to game-imposed randomness and constraints. This suggests that ``Overloading'' functions more as deception (hidden) rather than manipulation (forcing action). Conversely, ``Sycophant'' incorporates manipulative tactics, using stimuli to persuade players to take additional actions. This approach mirrors predatory advertising~\cite{petrovskaya2021predatory, Petrovskaya2022micro}, where rewards are used to create commercial ties with players, making them believe their actions are voluntary.

``Easy to Get, Hard to Lose'' parallels the ``Roach Motel'' pattern. Engaging games are easy to enter but difficult to leave, reinforcing psychological attachment. 
Similarly, \citet{petrovskaya2021predatory} identified a predatory monetization tactic where players receive a partial item for free but must pay to fully use it. We discovered a DP Combo ``Try Now and Pay Later,'' which required players to invest either money or time as effort, promising the full experience but taking it away from the player if they did not pay. This case shows how even small differences in similar designs can result in very different patterns of experience and potential harms.

The ``Never-Ending Lure'' pattern is designed to convert non-paying players into paying ones (``First Charge Discount'') and encourage consistent long-term payments (``Accumulating Rewards''). ``Accumulating Rewards'' employs a special manipulative strategy: players cannot directly purchase certain items, but they need a sufficient purchase history to be qualified to purchase them. \citet{petrovskaya2021predatory} identified predatory monetization strategies aligning with our findings, where players, feeling under-powered, were nudged to pay for significant in-game improvements with just a small amount of money.
In contrast to prior studies highlighting the unfair re-release of in-game products at lower prices~\cite{petrovskaya2021predatory, Petrovskaya2022micro}, ``Remedy Consumption'' operates differently by making re-released items more expensive and harder to obtain. This may be explained by the combination with ``Artificial Scarcity,'' positioning specific in-game products as luxury items, reflecting the strategic use of deceptive patterns to align with monetization goals.

Although these patterns could be implemented more ethically~\cite{xiao2022probability}, they often create exclusivity and premium experiences that favour paying players, pressuring non-paying players to invest money or time. These designs ultimately prioritize company interests, placing players in a vulnerable and unfair position.

\subsection{Deception and Dominance: Understanding the Distribution of Patterns}
Mobile gaming and free-to-play contexts may differ from other app domains~\cite{barr2007video}. In our corpora, we found differences in representation and dominance at all levels of the taxonomy.
The most prevalent general UI/UX DP we identified was ``Visual Prominence,'' which used a red dot beside buttons to repeatedly draw attention to new events and purchasable items, relying on the attention-guiding effect of the colour red~\cite{kuniecki2015color} to increase exposure to targeted marketing content~\cite{schmidt2015advertising}. Even in game UI/UX contexts, simple tweaks like a red beacon can be leveraged as DPs.

\textbf{Monetary} patterns were the most dominant, consistent with previous research. For example, ``Pay to Skip'' is common in casual mobile games~\cite{dahlan2022finding} and ``Pay Wall'' is characteristic of early childhood games~\cite{sousa2023dark}. However, we found few instances of ``Pay to Skip'' and no ``Pay Wall'' patterns. Instead, ``Intermediate Currency'' patterns and their subcategories were the most common. 
Perhaps our corpora did not include casual genres or target children. 
Similarly, while \citet{dahlan2022finding} identified ``Grinding'' as a dominant \textbf{Temporal} pattern, we found only one instance in a single game. However, we found widespread examples of other ``Playing by Appointment'' patterns across all corpora, albeit in different low-level forms. Like in \citet{sousa2023dark}, \textbf{Social} patterns were the least prevalent among all DPs. 
The \textbf{Psychological} patterns ``Complete the Collection'' and ``Aesthetic Manipulation'' were the most prevalent, aligning with \citet{sousa2023dark}. 
Rather than displaying what players have already achieved, these DPs emphasize the items that await players to unlock, which consequently pressures players to achieve ``All Collection'' as an ultimate goal~\cite{cruz2017need}. ``Aesthetic Manipulation'' uses visually appealing content to influence cognitive processes and emotions, further confirming the rich employment of psychological tricks.

\subsubsection{Formalizing the Notion of DP Combos}
Game developers often employ a combination of DPs in their designs~\cite{luguri2021shining}. 
Previous work has shown that the impact of a single DP can be amplified when combined with other DPs~\cite{baroni2021dark, mathur2019dark}. 
We found several cases in our game corpora. We formalize this approach as \textbf{DP Combos}. One possible explanation for the prevalence of DP Combos is the influence of the ``magic circle,'' where players are more tolerant of the suspension of social conventions and norms within the game world. This notion allows game developers greater freedom to design in-game content that diverges from real-world expectations. Consequently, developers can create a distinct and immersive game world where DPs are more readily accepted, further facilitating the use of DP Combos.

\subsubsection{Comparing the Chinese and Japanese Corpora}
In our exploratory analysis of the Chinese and Japanese corpora, we identified several patterns that seemed to be exclusive to each corpus. Given the sample size, we hesitate to draw firm conclusions, but offer our initial findings and suggest culture-orientated analyses for future work.

One key finding was the differing use of the ``\emph{Evil} Battle Pass.'' In the Chinese corpus, it was common to find multiple battle passes available simultaneously, each offering exclusive rewards. 
This tactic, which emphasizes the permanent loss of these rewards, propels the player to complete daily or weekly tasks and spend even more time in the game~\cite{hadan2024, frommel2022daily}. This pattern may relate to the rise of esports, with battle passes aligned not only with esports schedules but also events, enhancing audience interaction~\cite{yu2018game, pedra2024passive, joseph2021battle}. ``Remedy Consumption'' was widely used in the Chinese corpus, too. This DP, often combined with ``Artificial Scarcity,'' may be explained by the emerging trend of virtual luxury desires among young adults in China~\cite{sharma2024global, pedra2024passive}. 
``Never-Ending Lure'' was unique to the Chinese corpus. Duopoly in the Chinese mobile market~\cite{ling2023analysis} complicates the interpretation of our findings. These examples originated from the same developer, making it unclear whether they are game-specific, developer-specific, or represent a broader pattern. 
Another pattern specific to Chinese corpus was ``Friend Spam/Impersonation,'' where games can directly access the player's contact list. 
This may be attributed to high WeChat use in China~\cite{wang2019wechat}, where players can log into games directly through their social accounts. The prevalence of ``Reward Mania'' in the Chinese corpus may also relate to WeChat~\cite{wang2019wechat}, which offers a more convenient and direct marketing model~\cite{huang2021wechat} that enables games to easily connect with players. 

In the Japanese corpus, the use of ``Anchoring Tricks,'' ``FOMO,'' and ``Invested/Endowed Value'' was frequent, but were relatively rare in the Chinese corpus. While ``Anchoring Tricks'' shares a similar rationale with ``False Hierarchy,'' \citet{hidaka2023linguistic} identified the latter as the second most common DP in Japanese apps, indicating a preference for such DPs within Japanese development cultures. From a marketing perspective, FOMO is strongly correlated with collectivism~\cite{karimkhan2021fear}. Japanese culture, which is characterized by collectivist values~\cite{ogihara2017temporal}, may lead to FOMO having a significant impact on Japanese players. Unlike the match-based gameplay in the Chinese corpus, the Japanese corpus emphasized gradual player progression, as in the ``Invested/Endowed Value'' strategy. We also encountered the ``Alphabet Soup'' pattern, where foreign words are (mis)represented using local characters or symbols~\cite{hidaka2023linguistic}. We observed that in-game promotional offers were priced in ``USD,'' while other offers and purchasable items were listed in Japanese yen, the local currency. This design seems to be intentionally employed to manipulate and mislead players into purchasing these offers. Prices displayed in USD, like \$9.99, may appear more affordable compared to their equivalent in Japanese yen (1,500+ yen), particularly for players who are unfamiliar with the exchange rate.

\subsubsection{Complex and Evolving Deployments of DPs} 
Researchers have critiqued the issues associated with microtransactions~\cite{petrovskaya2021predatory, Petrovskaya2022micro}. 
Instead of simply marketing games for purchase, designers offer in-game items that can influence the game dynamics in a way that drives players to spend money~\cite{Petrovskaya2022micro}. 
Our findings reveal an important element: that game DPs tend to be overly complex and layered, with multiple deceptive and neutral mechanics in one place. To characterize this context, we introduced the concepts of ``DP Combos'' and ``DP Enhancers.'' \citet{hadan2024} identified the inclusion of multiple game DPs within the battle pass system of Overwatch 2. In our Chinese corpus, we not only confirmed this approach but also discovered that several games did so excessively. Although battle passes are typically designed to align with regular game seasons, developers have implemented parallel seasonal systems---such as esports competition seasons---to justify releasing multiple battle passes concurrently. 
This pushes players to purchase both, because advancing in either requires the same effort in-game. \citet{king2023investigating} identified the issue of using a virtual currency in predatory monetization with examples like Robux (Roblox's currency) and V-Bucks (Fortnite's currency). We found that not only are premium currencies commonly used across games, but the strategic use of multiple premium currencies within a single game was also prevalent. Developers design boundaries between these currencies to encourage players to make more purchases
; this seems to be intentional, as the currencies within one game are often non-interchangeable. This shows how certain DPs and approaches to their deployment are not only systematically employed but also evolve as developers play with different combinations and new patterns. 

\subsection{Surfacing DP Enhancers: From NPCs to Kawaii}
We identified certain elements that were used in combination with DPs to enhance their impact. Previous work has explored nudging as a form of persuasion in technology~\cite{hansen2013nudge, caraban201923ways}. While DPs are intended to deceive users, nudges are typically designed to improve user outcomes~\cite{ozdemir2020digital, sunstein2018better}. The \textbf{DP Enhancers} we identified do not deceive players directly but share similarities to nudge patterns by influencing the user's automatic or reflective decision-making processes.
This makes players more vulnerable to the DP/s accompanying the enhancer. These design choices are deeply rooted in the game context, gain significance from the interactions between players and NPCs, the rarity levels crafted by developers, and meaningful statistical values like ATK and DEF. For instance, \citet{king2023investigating} found that players were more likely to purchase items when requested or encouraged by NPCs. Similarly, we found that NPCs can enhance player engagement in gambling situations, such as loot boxes and lotteries, by encouraging the player to seek more powerful items or expressing pity when they fail. We call this DP Enhancer ``The Power of NPCs.''

Japanese cuteness or kawaii is widely used in Japanese role-playing games (JRPGs)~\cite{schules2015kawaii}, which have been exported to and are highly influential for the Chinese games industry~\cite{Kong_2024}. We observed cases where kawaii acted as an enhancer when combined with DPs like ``Sycophant,'' reducing player resistance to participating in out-of-game activities. Similarly, the common ``Rarity Level'' pattern, where in-game items are categorized into different rarity levels, was combined with ``Deceptive Luxury,'' which inflates the price of items to unreasonable levels. The ``Functional Value''  enhancer was often combined with ``Complete the Collection,'' providing statistical advantages to players who collected more characters, many of which could only be obtained through purchases, thus leading to unfair advantages between paying and non-paying players.

\subsection{Implications}
Our findings have several implications for academia, industry, and individual players.

\subsubsection{Examine DPs Across Interactive and Cultural 
Contexts}

Previous work has expanded our collective understanding of DPs by unearthing distinctions between game DPs and general varieties~\cite{zagal2013dark,sousa2023dark, dahlan2022finding}, discovering special DPs in specific cultural contexts, like Japan~\cite{hidaka2023linguistic}, identifying differences across platforms and devices~\cite{Gunawan2021}, 
and gathering insights from multiple stakeholders, such as experts and players~\cite{zagal2013dark,hadan2024,hadan2024comba,sousa2023dark,king2023investigating, dahlan2022finding}.
Our study builds on this body of work by focusing 
on providing an exploratory, descriptive analysis of DPs among Japanese and Chinese free-to-play mobile games. 
Our findings reinforce the importance of focusing on a certain interactive context, i.e., games, while demonstrating how general DPs cross UI/UX boundaries. We also observed similarities and differences in the types and distribution of game DPs compared to previous studies~\cite{dahlan2022finding, sousa2023dark}. 
We confirmed the prevalence of DPs in Chinese and Japanese free-to-play mobile games. However, the limited samples preclude broad generalizations across all Chinese and Japanese games, mobile and otherwise, so we could not contribute in-depth analyses from a sociocultural perspective. Future research should further investigate the sociocultural and sociolinguistic dimensions of DPs in Chinese, Japanese, and other cultural settings to elucidate how local factors shape the design, implementation, and experience of DPs.

\subsubsection{Actively Extend Ontologies of Deceptive Patterns}

We recognized the importance of developing an integrated ontology of game DPs to support future research on DPs generally and in games specifically. Certain game DPs share similar deceptive and manipulative characteristics with general DPs, while others are designed to coexist (DP Combos) or emerge alongside other patterns (DP Enhancers). To address this complexity, we constructed a three-level ontology, initially based on the categorization of general UI/UX DPs by \citet{gray2024ontology}. Our study extends the existing classification~\cite{zagal2013dark, sousa2023dark, king2023investigating} by identifying new game DPs 
and highlighting how these patterns deceive, mislead, and manipulate player behaviour both in the short- and long-terms. Future work should explore whether DP Enhancers and DP Combos exist in other interactive contexts and what effect these have on the user. Strategies could be developed using the ontology we extended here or other ontologies 
to help players recognize these patterns in practice.



\subsubsection{Advance Ethical Game Design}

Our findings indicate widespread use of game DPs, underscoring a pressing need for an ethical perspective on game design practice. Although DPs can be implemented without explicit intent~\cite{zhang-kennedy2024navigating, chivukula2018darkint}, we found that they are frequently employed to ``hook'' players and subtly lead them toward adopting microtransaction strategies. 
Specifically, we identified new game DPs premised on players spending real money in free-to-play games. The disparity between paying and non-paying players may pressure the latter to eventually spend money, a predatory tactic~\cite{petrovskaya2021predatory, Petrovskaya2022micro}. Such game DPs, which manipulate players over time, raise concerns about potential financial harm and problematic behaviours~\cite{xiao2022probability, bean2017video, zendle2020beyond, li2019relationship, hadan2024}. This highlights the importance of fostering company self-reflection, promoting ethical game design practice, and creating a fairer gaming environment. We recommend that game developers strike a balance between the interests of paying and non-paying players, creating a level playing field rather than encouraging payments to compete~\cite{freeman2022ingamepurch, petrovskaya2021predatory, Petrovskaya2022micro}. Future research should empirically assess the dynamics between multiple players in the context of free-to-play games and DPs, notably when players have financial disparities or differing needs.

\subsection{Limitations and Future Work}
\label{sec:limits}
We based our sample size and recording duration on previous work~\cite{sousa2023dark, dahlan2022finding, hochleitner2015heuristic}. We aimed to select representative games, but the diversity within commercial games may have led to gaps. DPs that may emerge later in gameplay were not captured. Future work can increase the sample size and number of sessions. Although we used the nine game genres, we did not compare DPs and distributions across these due to the sample size, which future research can explore. We focused on mobile games, which offer a relatively consistent standard for 
comparisons. Future research may examine free-to-play games on different platforms, such as PC or console games, for any platform-specific differences. 
The new DPs and DP designs (Combo and Enhancer) that we discovered should be investigated in other interactive and cultural settings, as their generalizability within and beyond games and the Chinese and Japanese markets remains unclear. We did not spend real-world money during gameplay, so we may have missed DPs related to in-game purchases. 
Also, we acknowledge that our ontology does not fully capture the complexity of certain DPs, like the ``\emph{Evil} Battle Pass.'' Future work should aim to develop an ontology that better reflects the multifaceted nature of individual DPs. Since this study focused exclusively on free-to-play games, future research should also examine pay-to-play games and conduct comparative analyses between the two models. Lastly, we did not consider the player's perspective. Future games UX research should examine actual experiences and attitudes toward the DPs explored and described here.

\section{Conclusion}
\label{sec:conclusion}
By analyzing free-to-play mobile games from Chinese and Japanese markets,  
we confirmed and expanded on the use and nature of deceptive game design patterns. We enriched established game-oriented DP ontologies by exploring and describing both existing and newly discovered classes and subclasses of game DPs within a novel setting. 
We hope this ontology will aid future research in this area. Additionally, we identified DP Combos and DP Enhancers, which illustrate how and when DPs are deployed in real gaming contexts. 
Efforts should be directed toward refining the framework and promoting ethical game design practice, ultimately benefiting academia, industry, and individual players.

\section*{Note on Machine Translation and Generative AI Use:} We used ChatGPT and Grammarly to correct grammar and ensure formal academic phrasing. The original draft was primarily written in English, but some parts were machine-translated using ChatGPT from Mandarin Chinese; we have retained the original text for comparison.

\begin{acks}
We thank Hiroki Kamakura for assisting Taro Nakajima with Japanese translation and direction during the recording of the gameplay. Illustrative examples of DPs presented in this paper were developed using design resources from \href{www.freepik.com}{Freepik}. This work was funded in part by a Japan Science and Technology Agency (JST) PRESTO grant (\#JPMJPR24I6).
\end{acks}

\bibliographystyle{ACM-Reference-Format}
\bibliography{REFS}

\end{document}